\documentclass[fleqn,usenatbib]{mnras}
\usepackage[T1]{fontenc}
\usepackage{ae,aecompl}
\usepackage{graphicx}	
\usepackage{amsmath}	
\usepackage{amssymb}	
\usepackage{lscape}
\newcommand{\HII}{H\small{II}\normalsize{}}
\defcitealias{bruzual03}{BC03}

\title[CIGALE for Star Clusters]{PHANGS--HST: Star Cluster Spectral Energy Distribution Fitting with CIGALE}

\author[J. A. Turner et al.]{
Jordan~A.~Turner,$^{1}$\thanks{E-mail: jturne19@uwyo.edu}
Daniel~A.~Dale,$^{1}$
Janice~C.~Lee,$^{2}$
M\'ed\'eric~Boquien,$^{3}$
Rupali~Chandar,$^{4}$
\newauthor
Sinan~Deger,$^{2}$
Kirsten~L.~Larson,$^{2}$
Angus~Mok,$^{4}$
David~A.~Thilker,$^{5}$
Leonardo~Ubeda,$^{6}$
\newauthor
Bradley~C.~Whitmore,$^{6}$
Francesco~Belfiore,$^{7}$
Frank~Bigiel,$^{8}$
Guillermo~A.~Blanc,$^{9,10}$
Eric~Emsellem,$^{11,12}$
\newauthor
Kathryn~Grasha,$^{13}$
Brent~Groves,$^{13}$
Ralf~S.~Klessen,$^{14,15}$
Kathryn~Kreckel,$^{16}$
J.~M.~Diederik~Kruijssen,$^{16}$
\newauthor
Adam~K.~Leroy,$^{17}$
Erik~Rosolowsky,$^{18}$
Patricia~Sanchez-Blazquez,$^{19,20}$
Eva~Schinnerer,$^{21}$
\newauthor
Andreas~Schruba,$^{22}$
Schuyler~D.~Van~Dyk,$^{2}$
and Thomas~G.~Williams$^{21}$
\\
$^{1}$Department of Physics \& Astronomy, University of Wyoming, Laramie, WY USA\\
$^{2}$Caltech/IPAC, Pasadena, CA USA\\
$^{3}$Centro de Astronom\'ia (CITEVA), Universidad de Antofagasta, Antofagasta, Chile\\
$^{4}$Department of Physics \& Astronomy, University of Toledo, Toledo, OH USA\\
$^{5}$Center for Astrophysical Sciences, The Johns Hopkins University, Baltimore, MD USA\\
$^{6}$Space Telescope Science Institute, Baltimore, MD USA
$^{7}$INAF -- Osservatorio Astrofisico di Arcetri, Largo E. Fermi 5, I-50157, Firenze, Italy\\
$^{8}$Argelander-Institut f\"ur Astronomie, Universit\"at Bonn, Auf dem Hügel 71, D-53121 Bonn, Germany\\
$^{9}$The Observatories of the Carnegie Institution for Science, 813 Santa Barbara St., Pasadena, CA, 91101\\
$^{10}$Departamento de Astronom\'ia, Universidad de Chile, Camino del Observatorio 1515, Las Condes, Santiago, Chile\\
$^{11}$European Southern Observatory, Karl-Schwarzchild Stra\ss e 2, D-85748 Garching bei M\"unchen, Germany\\
$^{12}$Universit\'e Lyon 1, ENS de Lyon, CNRS, Centre de Recherche Astrophysique de Lyon UMR5574, F-69230 Saint-Genis-Laval, France\\
$^{13}$Research School of Astronomy and Astrophysics, Australian National University, Canberra ACT, Australia\\
$^{14}$Universit\"at Heidelberg, Zentrum f\"ur Astronomie, Institut f\"ur Theoretische Astrophysik, Heidelberg, Germany\\
$^{15}$Universit\"at Heidelberg, Interdisziplin\"ares Zentrum f\"ur Wissenschaftliches Rechnen, Heidelberg, Germany\\
$^{16}$Astronomisches Rechen-Institut, Zentrum f\"{u}r Astronomie der Universit\"{a}t Heidelberg, Heidelberg, Germany\\
$^{17}$Department of Astronomy, The Ohio State University, Columbus, Ohio 43210, USA\\
$^{18}$Department of Physics, University of Alberta, Edmonton, AB, Canada\\
$^{19}$Departamento de F\'isica de la Tierra y Astrof\'sica, Universidad Complutense de Madrid, E-28040, Spain\\
$^{20}$Instituo de F\'isica de Particulas y del Cosmos IPARCOS, Universidad Complutense de Madrid, E-28040, Madrid, Spain\\
$^{21}$Max-Planck-Institut f\"ur Astronomie, K\"onigstuhl 17, D-69117 Heidelberg, Germany\\
$^{22}$Max-Planck-Institut f\"ur Extraterrestrische Physik, Giessenbachstra{\ss}e 1, D-85748 Garching, Germany
}

\date{Accepted XXX. Received YYY; in original form ZZZ}

\pubyear{2020}

\begin{document}
\label{firstpage}
\pagerange{\pageref{firstpage}--\pageref{lastpage}}
\maketitle

\begin{abstract}
The sensitivity and angular resolution of photometric surveys executed by the {\it Hubble Space Telescope} (\textit{HST}) enable studies of individual star clusters in galaxies out to a few tens of megaparsecs. The fitting of spectral energy distributions (SEDs) of star clusters is essential for measuring their physical properties and studying their evolution. We report on the use of the publicly available Code Investigating GALaxy Emission (\textsc{cigale}) SED fitting package to derive ages, stellar masses, and reddenings for star clusters identified in the Physics at High Angular resolution in Nearby GalaxieS--HST (PHANGS--HST) survey. Using samples of star clusters in the galaxy NGC~3351, we present results of benchmark analyses performed to validate the code and a comparison to SED fitting results from the Legacy ExtraGalactic Ultraviolet Survey (LEGUS). We consider procedures for the PHANGS--HST SED fitting pipeline, e.g., the choice of single stellar population models, the treatment of nebular emission and dust, and the use of fluxes versus magnitudes for the SED fitting. We report on the properties of clusters in NGC~3351 and find, on average, the clusters residing in the inner star-forming ring of NGC~3351 are young ($<10$~Myr) and massive ($10^{5}~M_{\odot}$) while clusters in the stellar bulge are significantly older. Cluster mass function fits yield $\beta$ values around $-2$, consistent with prior results with a tendency to be shallower at the youngest ages. Finally, we explore a Bayesian analysis with additional physically-motivated priors for the distribution of ages and masses and analyze the resulting cluster distributions. 
\end{abstract}


\section{Introduction}
The measurement of the physical properties of stellar populations using broadband photometry is an arduous task. Even at the level of an individual single-aged star cluster, various phenomena are at play. As starlight interacts with dust, a fraction is absorbed and then re-emitted at longer wavelengths. Additionally, the dust obscures the view of the star clusters leading to extinction and reddening of the stellar emission. Young clusters contain massive stars that ionize the surrounding gas, leading both to the rise of a nebular continuum and to the appearance of a series of emission lines. This nebular emission can be quite bright, and may represent a non-negligible fraction of the flux captured by broadband filters \citep[e.g.,][]{anders03,groves08,boquien10,reines10}. In addition to the effects from dust and ionized gas, there are dependencies on the initial mass function (IMF) of the stellar population where the IMF can be fully-sampled or stochastically sampled \citep[e.g.,][]{barbaro77,girardi93}. To further our understanding of star cluster formation and evolution, it is necessary to carefully account for these effects to accurately measure fundamental cluster properties such as the age, mass, and reddening.

These measurements of stellar cluster properties provide an essential tool for understanding the mechanisms which drive, regulate, and extinguish star formation at small scales. In turn, catalogs of star clusters can be combined with observations of gas in nearby galaxies to chart the cycling of gas into stars, allowing us to study the dependence of star formation on environmental parameters on galactic scales. Previously developed catalogs of stellar cluster ages and masses have enabled the study of cluster mass and age functions \citep[e.g.,][]{fouesneau14,ashworth17,chandar17,linden17,mok19}, star formation in different environments \citep[e.g.,][]{whitmore14,chandar17,leroy18}, and star formation efficiencies and timescales \citep[e.g.,][]{grasha18,grasha19}. Most recently, the Legacy ExtraGalactic Ultraviolet Survey (LEGUS) project \citep{calzetti15} has derived the ages, masses, and reddenings of star clusters using the SED modelling techniques described in \cite{adamo17}.

Over the past decades, various codes have been developed to model the panchromatic emission from galaxies. Combined with Bayesian techniques, they can provide deeper insight into the properties of star cluster populations in conjunction with simple $\chi^2$ minimization for constraining physical parameters \citep[e.g.,][]{dacunha08,franzetti08,han12,moustakas13,chevallard16}. Although this machinery has enabled significant advances for the interpretation of galaxy spectral energy distributions (SEDs), application to stellar clusters has been more limited. 

The availability of public codes under permissive licenses offers an opportunity to bring the benefits enjoyed by galaxy studies to studies of stellar clusters. Here, we report on an augmentation of the publicly available SED fitting package \textsc{cigale}: \textit{Code Investigating GALaxy Emission}\footnote{\url{https://cigale.lam.fr}} \citep{burgarella05,noll09,boquien19} to derive ages, stellar masses, and reddenings for star clusters identified in the Physics at High Angular resolution in Nearby GalaxieS--HST (PHANGS--HST) survey (J.\ C.\ Lee et al. in prep.)\footnote{\url{https://phangs.stsci.edu}}. PHANGS\footnote{\url{http://www.phangs.org}} is a panchromatic collaboration comprised of: PHANGS--ALMA, a large \mbox{CO(2--1)} mapping program aimed at covering a representative sample of $\sim$74 nearby galaxies (A.\ K.\ Leroy et al., in prep.); PHANGS--MUSE, a Very Large Telescope (VLT) imaging program of $\sim$20 of the PHANGS galaxies with the MUSE optical IFU instrument \citep[E.\ Emsellem et al. in prep.; see first results in][]{kreckel16,kreckel19}; and PHANGS--HST which aims to chart the connections between molecular clouds and young star clusters/associations throughout a range of galactic environments by imaging the 38 galaxies from the PHANGS sample best-suited{\footnote{Galaxies that are relatively face-on, avoid the Galactic plane, and have robust molecular cloud populations to facilitate joint analysis of resolved stellar populations and molecular clouds (G.\ S.\ Anand et al. in prep; J.\ C.\ Lee et al. in prep.).}} for study of resolved stellar populations. PHANGS--HST observations, which begun in 2019 April and are scheduled to conclude in mid 2021, are expected to yield $NUV$-$U$-$B$-$V$-$I$ photometry for tens of thousands of stellar clusters and associations. 

\textsc{cigale} is designed for speed, ease of use, and adaptability. It is based on a Bayesian approach for estimating physical properties and the corresponding uncertainties. It has the necessary flexibility for handling different stellar evolution tracks, star formation histories, dust attenuation curves, and options for including nebular emission. A $\chi^2$ minimization option is available with \textsc{cigale} to simplify comparisons with prior work. It is well-suited for supporting stellar cluster studies, as well as for self-consistent modelling of both single-aged and composite stellar populations. Such an approach is needed for the characterization of structures across the full star formation hierarchy, beyond the densest peaks (i.e., stellar clusters), as will be investigated by PHANGS. 

For this specific work, \textsc{cigale} has been been expanded to handle the modelling of pure single-aged stellar populations, and the \textsc{yggdrasil} stellar populations models \citep{zackrisson11} have been added to allow for easy comparison with previous studies, complementing the \citet[][hereafter \citetalias{bruzual03}]{bruzual03} populations that are provided by default.\footnote{There are a number of star formation history (SFH) and single stellar population (SSP) modules available by default within \textsc{cigale}. The SFH modules included are a double exponential, a delayed SFH with optional exponential burst or constant burst/quench, a periodic SFH, and a user specified SFH read in from an input file. The SSP models included by default are \cite{bruzual03} and \cite{maraston05}.} \textsc{cigale} fits photometry in linear flux units by default, but a modification to enable the fitting of photometry in units of magnitudes has also been implemented to allow for easy comparison with previous star cluster SED modelling studies. These changes are available in dedicated branches, \texttt{SSP} and \texttt{SSPmag}, respectively, of the public \texttt{git} repository\footnote{\url{https://gitlab.lam.fr/cigale/cigale.git}} of \textsc{cigale}. When employing these modified versions, one still has the option to run \textsc{cigale} in its default operating mode of modelling the SFH rather than to fit the SSP track directly.

This paper is part of a series which documents the major components of the overall PHANGS--HST data products pipeline: survey design and implementation (J.\ C.\ Lee et al. in prep.); source detection and selection of compact star cluster candidates (D.\ A.\ Thilker et al. in prep.); aperture correction and quantitative morphologies of star clusters (S.\ De\u{g}er et al. in prep.); star cluster candidate classification (B.\ C.\ Whitmore et al. in prep.); neural network classification proof-of-concept demonstration \citep{wei20}; stellar association identification and analysis (K.\ L.\ Larson et al. in prep.); and constraints on galaxy distances through analysis of the Tip of the Red Giant Branch (TRGB) as observed in the PHANGS--HST parallel pointings \citep{anand20}. 

Here, we focus on the methodology for fitting the $NUV$-$U$-$B$-$V$-$I$ photometry for star clusters with \textsc{cigale}. In Section~\ref{sec:data}, we review the data utilized in this work. In Section~\ref{sec:benchmark}, we present the results of a benchmark analysis to validate the code.\footnote{See also \cite{hunt19} for a recent benchmark study involving \textsc{cigale} for the modelling of galaxy SEDs.} We use \textsc{cigale} to fit mock cluster photometry and examine the accuracy of the recovered properties. We also use \textsc{cigale} to fit photometry for stellar clusters in NGC~3351 published by the LEGUS project, and compare with results from their proprietary stellar cluster SED fitting code \citep{adamo17}. In Section~\ref{sec:physical_properties}, we use new imaging data obtained by PHANGS--HST for a larger region of NGC~3351 to help establish procedures for the PHANGS--HST SED fitting pipeline using \textsc{cigale} (e.g., single stellar population models to be adopted, treatment of nebular emission and dust), and quantify model dependencies in the results. In Section~\ref{sec:results}, we take an initial look at the stellar cluster age, mass, and reddening results for NGC~3351 and explore spatial dependencies as well as the mass functions. Additionally, we explore the application of physically-motivated Bayesian priors. Finally, in Section~\ref{sec:future}, we discuss possible future additions and tweaks to the SED modelling pipeline. We end by summarizing our findings in Section~\ref{sec:conclusions}.

\begin{figure}
	\includegraphics[width=\columnwidth]{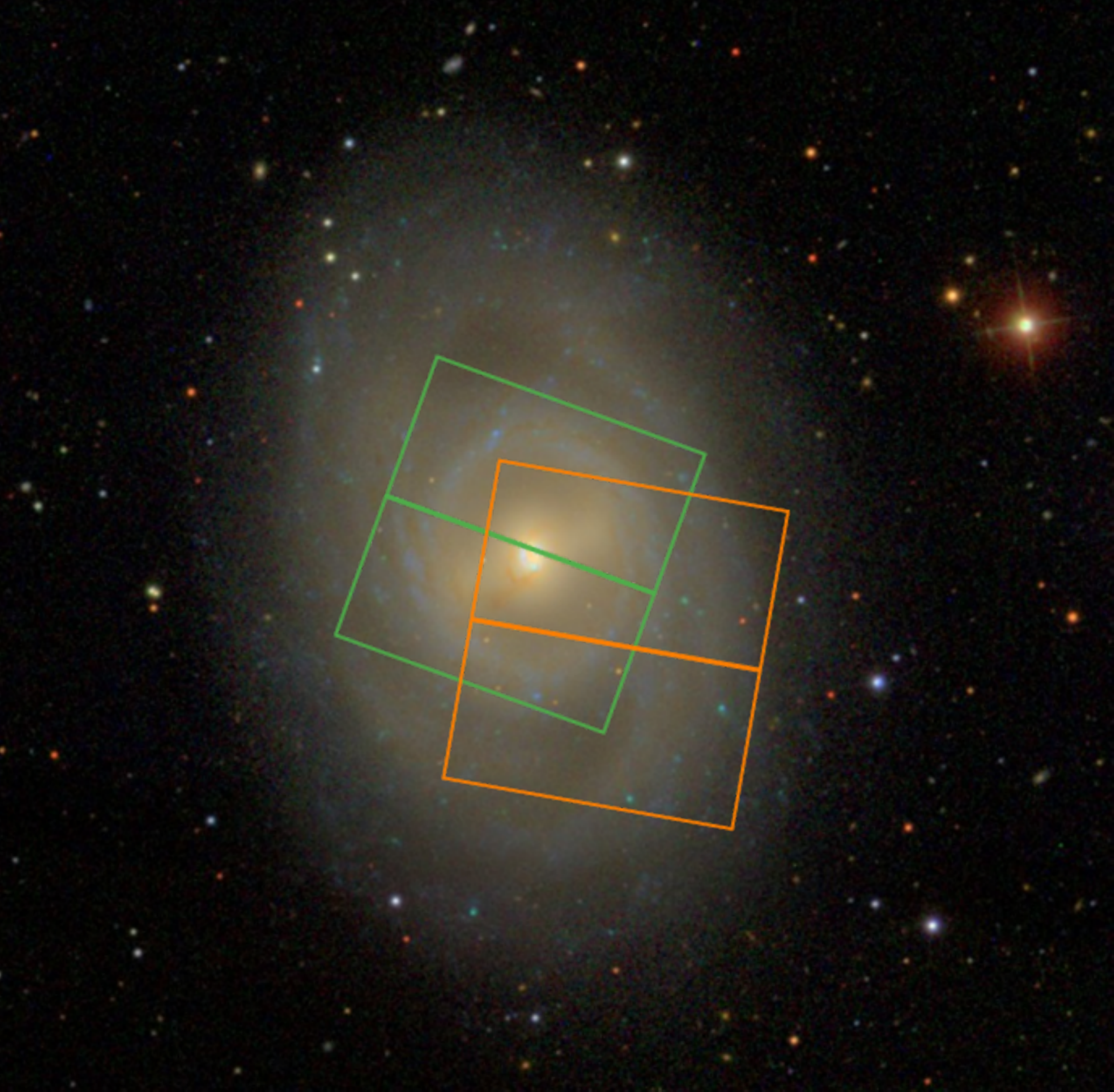}
    \caption{WFC3 observation footprints from PHANGS--HST (green) and LEGUS (orange) overlaid on a Sloan Digital Sky survey $g$-$r$-$i$ image of NGC~3351 (David W. Hogg, Michael R. Blanton, and the Sloan Digital Sky Survey Collaboration). The WFC3 field of view is 162$\arcsec\times$162$\arcsec$.}
    \label{fig:footprints}
\end{figure}

\begin{table}
\centerline{Table 1. $HST$ Integrations for NGC~3351}
\begin{tabular}{lcccccccc}
\hline
\hline

Name & F275W & F336W & F438W & F555W & F814W & \\
 & [s] & [s] & [s] & [s] & [s] \\
\hline

PHANGS		& 2190 & 1110 &       1050  &   ~670  & ~830 \\
LEGUS		& 2361 & 1062 &       ~908  &   1062  & ~908 \\
\hline
\end{tabular}
\caption{\textit{HST} WFC3 UVIS imaging exposure times corresponding to the footprints illustrated in Figure~\ref{fig:footprints}. New data for NGC~3351 were obtained by PHANGS--HST in all five filters, in order to better cover the region mapped in CO with ALMA.}
\label{tab:observations}
\end{table}

\section{Data}
\label{sec:data}

\subsection{NGC~3351 HST Imaging}
To exercise our SED-fitting procedures, we use photometry of stellar clusters measured from {\it Hubble Space Telescope} (\textit{HST}) $NUV$-$U$-$B$-$V$-$I$ (F275W-F336W-F438W-F555W-F814W) imaging of the nearby galaxy NGC~3351. NGC~3351 is an Sb spiral at 10.0~Mpc \citep{freedman01} with an approximately solar metallicity \citep{moustakas10}. Measurements using PHANGS--MUSE H\,\textsc{ii} regions confirm this, and also show a flat radial metallicity gradient across our field of view (K.\ Kreckel priv.\ comm.), which simplifies the interpretation of results based upon the standard assumption of a single metallicity model for the entire cluster population of a galaxy.

Though 80 per cent of the PHANGS--HST galaxy sample had no existing \textit{HST} wide-field imaging prior to the start of the program, NGC~3351 is one of the exceptions. It was observed by LEGUS in 2014 and was one of the first targets to be observed in the PHANGS program in 2019 May. While LEGUS observations were taken to maximize radial coverage of the galaxy including the nucleus, PHANGS--HST seeks to maximize the coverage of available PHANGS--ALMA \mbox{CO(2--1)} mapping, which leads to overlapping but complementary observations as shown in Figure~\ref{fig:footprints}. Exposure times are given in Table~\ref{tab:observations}. Star cluster photometry along with masses, ages, and reddening from SED fitting for sources detected in the area observed by LEGUS \citep{adamo17} are publicly available through MAST from LEGUS.\footnote{\url{https://legus.stsci.edu}} Not only does this make NGC~3351 an excellent choice for a benchmark analysis, but the galaxy has been well studied due to its circumnuclear star formation \citep[e.g.,][]{buta88,debbie97}, and it is also in the SINGS sample \citep{kennicutt03}. Hence, there is a wealth of ancillary data available, which provide independent constraints on parameters of interest here, such as dust reddening and metallicity, albeit at lower resolutions.

NGC~3351 is also an interesting subject for a first look at the variation of cluster properties in different environments using the PHANGS--HST dataset, again due to its morphological and dynamical structure. This barred spiral galaxy exhibits a large range in optical surface brightness and levels of obscuration by dust between its inner and outer star-forming rings (the latter of which is now completely sampled with the addition of new PHANGS--HST imaging as shown in Figure~\ref{fig:footprints}).

\subsection{NGC3351 Star Cluster Photometry}

For our SED fitting analysis, we use $NUV$-$U$-$B$-$V$-$I$ photometry of star clusters which have been identified by the PHANGS--HST pipeline, visually inspected and classified. The PHANGS--HST methodology for source detection, candidate selection, and cluster classification builds upon the process developed by LEGUS. The LEGUS procedure is described in detail in \citet{grasha15} and \citet{adamo17}, and we give a brief overview here. Source detection was performed with \textsc{SExtractor} on the $V$-band image. Clusters selected for analysis and whose properties were derived via SED fitting were those objects which: (1) had a concentration index (CI) greater than 1.3 (i.e., a difference in magnitudes between circular apertures with radii of 3 and 1 pixels, which indicates that the object is more extended than a point source); (2) had $M_{V}<-6$ Vega mags (after aperture correction) and were detected in at least 4 of 5 filters; (3) had been visually inspected and classified as either a class~1: symmetric compact cluster, class~2: asymmetric compact cluster, class~3: multi-peaked compact association.

In the PHANGS--HST and LEGUS footprints, we find a total of 468 clusters (with 133, 166, 166 visually inspected to be class~1, class~2, and class~3 objects using the same classification criteria as LEGUS). Of these, 136 are within the new area of NGC3351 observed by PHANGS--HST (with 18, 56, and 61 visually inspected to be class~1, class~2, and class~3 objects). Details of the PHANGS--HST cluster identification methodology is presented in D.\ A.\ Thilker et al. (in prep.) and B.\ C.\ Whitmore et al. (in prep.).
 
A detailed comparative analysis of the PHANGS--HST and LEGUS clusters catalogs is presented in D.\ A.\ Thilker et al. (in prep.), and comparison with the stellar association catalog is presented in K.\ L.\ Larson et al. (in prep.). Overall, the PHANGS--HST and LEGUS catalogs for the imaging observations originally obtained by LEGUS for NGC~3351 contain a comparable number of clusters to $M_{V}<-6$ Vega magnitude (after aperture correction) with $\sim$75 per cent overlap.

Photometry is performed on each cluster with an aperture radius of 4 pixels which, at the distance of NGC~3351, corresponds to a physical scale of $7.7$~pc. To account for extended emission beyond the 4 pixel radius, an aperture correction of 0.68~mag in the $V$-band is applied ($NUV$: 0.87~mag, $U$: 0.80~mag, $B$: 0.71~mag, $I$: 0.80~mag) independent of cluster profile. Foreground extinction to NGC~3351 due to the Milky Way is computed following \cite{schlafly11}, which adopt \cite{schlegel98} reddening maps and the \cite{fitzpatrick99} reddening law with $R_{V} = 3.1$. The details of the aperture correction derivation is presented in S.\ De\u{g}er et al. (in prep.). The median signal-to-noise ratio for a cluster in the $V$-band is $\sim$45 (min: $\sim$8, max: $\sim$310).

\section{Benchmark Testing \textsc{cigale}}
\label{sec:benchmark}

\subsection{CIGALE: Basic Considerations}
\label{subsec:cigale}
\textsc{cigale} operates by generating a grid of models based on the user's input parameters. In our case, the model grid samples two free parameters: age and reddening based on our chosen single-age stellar population (SSP) model, e.g., the \citetalias{bruzual03} models. The sampling of the age and reddening grids is chosen by the user. For the age grid, we have ten linearly spaced models for 1 to 10~Myr ($\Delta T=$~1~Myr, the highest precision available to \textsc{cigale}) and 100 evenly log-spaced models for 11 to 13\,750~Myr ($\Delta\log(T/{\rm Myr})\approx0.3$). Although cluster mass is an output of the SED modeling, it is not treated as a third dimension of the model grid. The masses corresponding to a particular model on the age-reddening grid are determined directly from the chosen IMF and star formation history. The masses are normalized to 1~$M_{\odot}$ at birth and once \textsc{cigale} fits a cluster's SED based on the age and reddening, the mass is appropriately scaled based on the cluster's luminosity. A fully-sampled IMF is assumed. The effect of a stochastically sampled IMF is discussed in Section~\ref{sec:stoch}. Model assumptions are listed in Table~\ref{tab:parameters}\ref{tab:params_phangs}.

\textsc{cigale} compares the cluster's photometry with each model of the grid and calculates the $\chi^{2}$ value to determine the goodness-of-fit. The $\chi^{2}$ value is converted into a likelihood via $\exp(-\chi^{2}/2)$. Once each model has been tested, \textsc{cigale} estimates the best-fit parameters in two ways: simple $\chi^{2}$ minimization and a likelihood-weighted mean \citep[see][section 4.3]{boquien19}. The model with the lowest $\chi^{2}$ is the `best-fitting' result and allows for easy comparison with other SED fitting procedures commonly used in the past \citep[e.g.,][]{chandar10b,adamo17}. $1\sigma$ uncertainties can be calculated by the difference between the best-fitting model and the models with $\chi^2_{\rm reduced}$ values of $1 + \chi^2_{\rm reduced, min}$. The likelihood-weighted mean of all the models on the grid is computed which is used as a Bayesian estimate for the physical properties. $1\sigma$ uncertainties are determined by the likelihood-weighted standard deviation of all the models.\footnote{In other words, we calculate the likelihood-weighted mean of the marginalized probability distribution function (PDF) with only a flat, bounded prior. The $1\sigma$ uncertainties describe the width of the peak of the PDF.} Note that while the figures of this paper present the ages and masses in logarithmic units, \textsc{cigale} performs all analyses in linear units.

For the first part of our analysis we examine results based on the best-fitting ($\chi^{2}$ minimized) values, which facilitates comparison to the body of previous star cluster work. Later in Section~\ref{sec:prior}, we explore differences when additional Bayesian priors are imposed on the age and mass distributions and how to estimate the cluster properties from the posterior probability distribution functions.

\subsection{Recovery of Mock Clusters}

\begin{figure*}
    \centering
    \includegraphics[width=\textwidth]{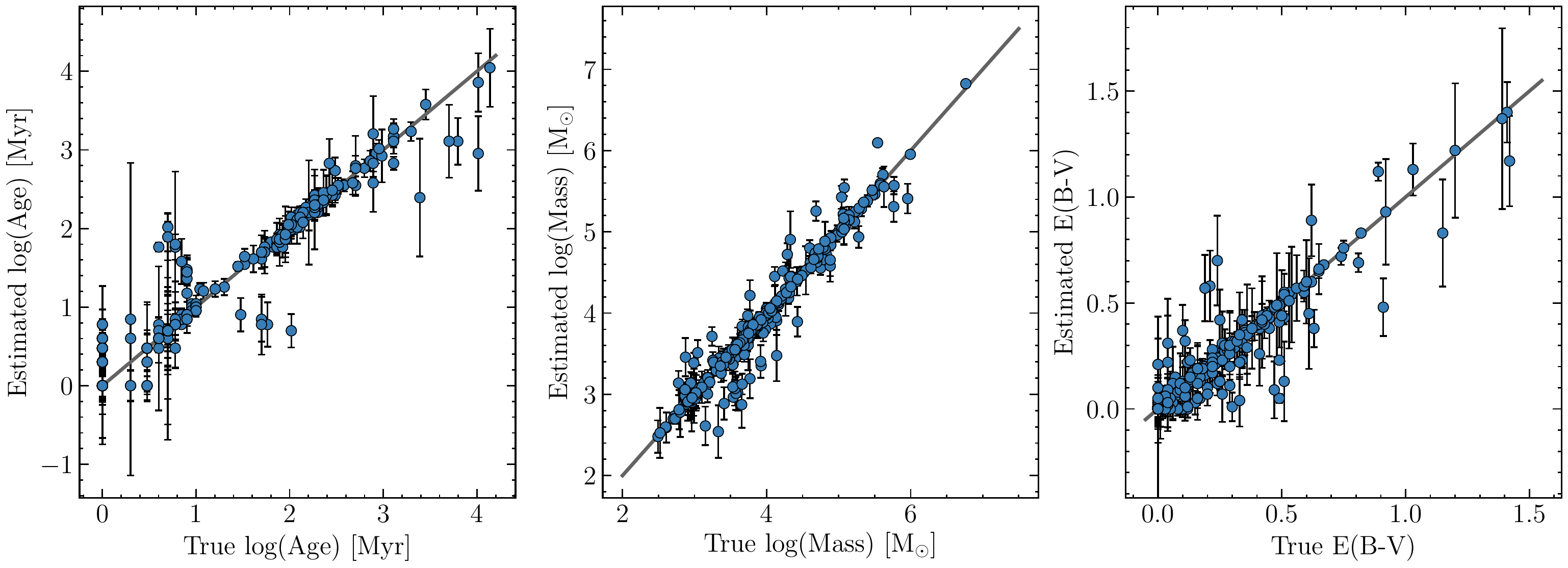}
    \caption{\textsc{cigale} best-fitting values compared to the `true' input values. $1\sigma$ error bars are given for each cluster's estimated property and account for most of the scatter about unity. Cluster ages are well recovered across the sample except for clusters with `true' ages at 1~Myr (0.25~dex scatter), at around 10~Myr (0.34~dex scatter), and at the oldest ages (0.39~dex scatter). Both the masses and reddenings are well recovered across the sample.}
    \label{fig:mock}
\end{figure*}

First, to determine how well cluster properties can be constrained, we generate mock cluster photometry for known cluster ages, masses, and reddenings to estimate how well \textsc{cigale} recovers these cluster properties. As our starting point, we adopt the 296 model SEDs (i.e., model $NUV$, $U$, $B$, $V$, $I$ fluxes, ages, masses, and reddenings) that best fit ($\chi^{2}$ minimized) the PHANGS--HST visually-classified class~1 and 2 star clusters in NGC~3351 which have photometric detections in all five bands. The mock cluster fluxes are then produced by randomly selecting a flux from a Gaussian distribution centered on the model flux with a standard deviation based on the median photometric uncertainty in each band of our PHANGS--HST cluster catalog. The median uncertainties (for photometry within a 4~pixel radius aperture; see D.\ A.\ Thilker et al. in prep.) are 5.02 per cent for F275W, 5.12 per cent for F336W, 3.51 per cent for F438W, 2.24 per cent for F555W, and 3.21 per cent for F814W. We then run \textsc{cigale} on this mock catalog and compare the resulting best-fitting values with the `true' input values as shown in Figure~\ref{fig:mock}. 

\begin{figure}
    \centering
    \includegraphics[width=\columnwidth]{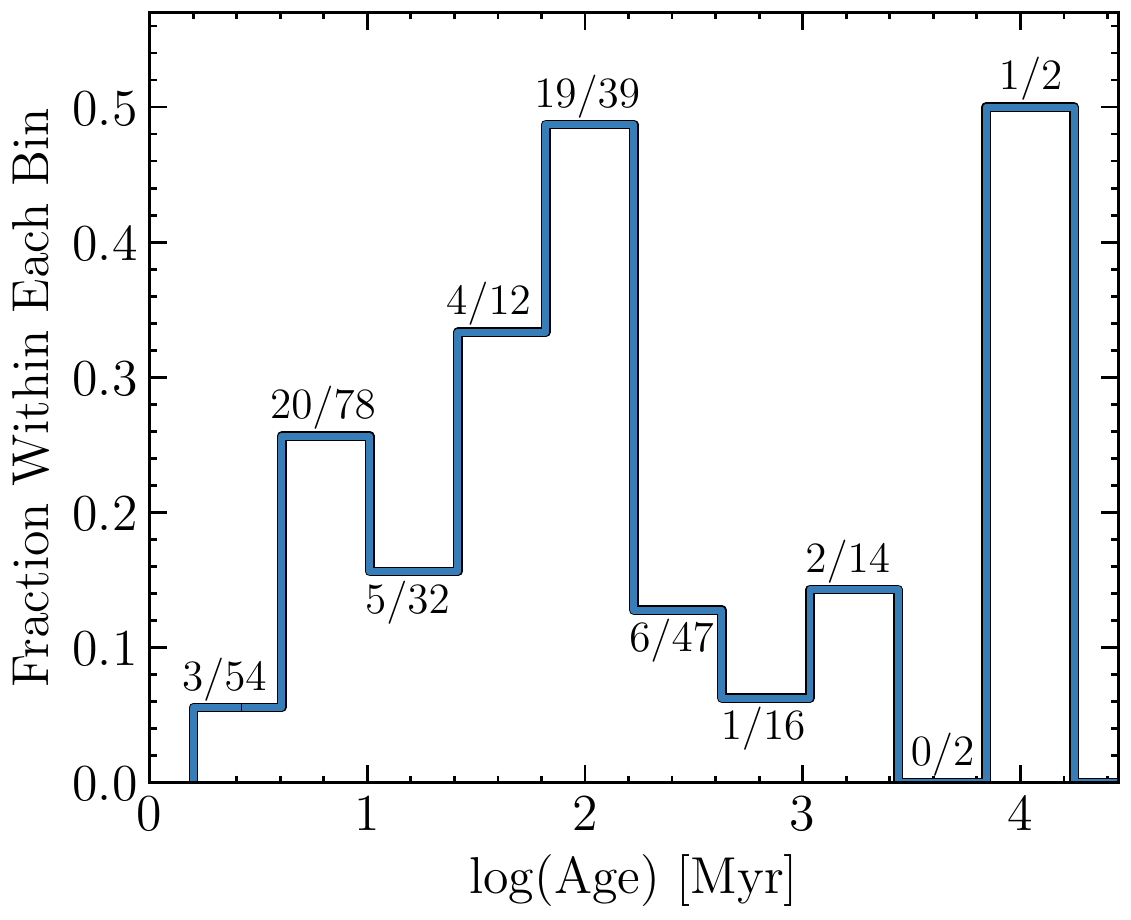}
    \caption{Histogram of the recovered ages of the bimodal cases as a fraction of the total number of mock clusters within each bin. The fraction of bimodal cases to the total number clusters is given at the top of each bin. The majority of the bimodal cases are concentrated at around 10~Myr and 100~Myr.}
    \label{fig:mock_hist}
\end{figure}

As Figure~\ref{fig:mock} shows, cluster ages are well recovered across the sample within the uncertainties except at 1~Myr, at just below 10~Myr, and for the oldest clusters. From 10~Myr to $\sim$30~Myr, the evolutionary tracks loop back onto themselves (see Figure~\ref{fig:color_color_tracks}) which introduces degeneracies of the available SEDs which map to the SSP models at those ages. Across the entire sample, the standard deviation of the difference between the `true' and recovered ages, masses, and reddenings are 0.31~dex, 0.18~dex, and 0.09~mag, respectively. There is no significant systematic offset between the true and recovered ages. 

The clusters with large residuals are found to be mostly those with probability distribution functions (PDFs) which exhibit bimodality where the young models with high reddening and older models with less reddening are both likely. We find $\sim$20 per cent of the estimated cluster sample to have bimodal PDFs. Figure~\ref{fig:mock_hist} shows the distribution of recovered ages for the bimodal cases as a fraction of the total number of mock clusters across ten age bins. The grouping of clusters with true ages around 100~Myr and underestimated age results have PDFs which are all bimodal. In this case, the $\chi^{2}$ minimization chose the young mode when the old mode was the true age. A majority of the clusters with true ages at 10~Myr and overestimated age results are bimodal as well. In this case, the opposite is true; the $\chi^{2}$ minimization chose the old mode when the young mode was the true age. The bimodality causes an increase in the scatter for the estimated $E(B{-}V)$ versus the true $E(B{-}V)$ values. Aside from these cases, for the most part, the reddenings are recovered well with a median difference from the `true' reddening value of 0.0~mag with a dispersion of 0.09~mag. The cluster masses are also recovered well across the mock catalog sample within the uncertainties. The median logarithmic mass ratio is $0.01$~dex with a dispersion of 0.18~dex. Overall, we can expect \textsc{cigale} to behave consistently within our photometric uncertainties with the exception of the bimodal cases which are discussed in depth in later sections.

\subsection{Comparison with LEGUS SED Modeling}

As a second method of benchmark testing, we use the published photometry\footnote{\url{https://archive.stsci.edu/prepds/legus/cluster_catalogs/ngc3351.html}} for the 292 clusters identified by LEGUS in their \textit{HST} imaging of NGC~3351, fit the photometry with \textsc{cigale}, and compare to results from the SED modeling performed by LEGUS as described in \cite{adamo17}. For the benchmark testing, we focus only on the 289 LEGUS clusters with photometric measurements in all five bands. We adopt the same assumptions used to produce the LEGUS ``reference'' catalogs \citep{adamo17}; Table~\ref{tab:parameters}\ref{tab:params_legus} summarizes the parameters adopted for the benchmark comparison. Briefly, \cite{adamo17} utilized the Padova--AGB stellar evolution isochrones and the \textsc{yggdrasil} population synthesis code \citep{zackrisson11} to generate single-aged stellar population models. The LEGUS work assumes a \cite{kroupa01} IMF from $0.1-100~M_{\odot}$; the \cite{cardelli89} Milky Way extinction law; flux from the nebular continuum and emission lines with a fixed covering fraction of 0.5; and solar metallicity isochrones. The minimum reddening is $E(B{-}V)=0$~mag and the maximum is set to be $E(B{-}V)=1.5$~mag with steps of 0.01~mag. The models are reddened before being fitted to the observed photometry. 

\makeatletter
\newcommand\newtag[2]{#1\def\@currentlabel{#1}\label{#2}}
\makeatother

\begin{table*}
\centerline{Table 2. SED Fit Choices}
\begin{tabular}{ll}
\hline \hline
\newtag{A}{tab:params_legus}) LEGUS Benchmark Testing\\
\hline
Star formation history          & instantaneous burst\\
Reddening \& Extinction         & $E(B{-}V)=$~[0\,:\,1.5]~mag; $\Delta=0.01$~mag; $R_V=A_V/E(B{-}V)=3.1$\\
SSP model                       & Yggdrasil Padova-AGB \\
Metallicity                     & $Z=0.02$ (i.e., solar metallicity)\\
IMF                             & Kroupa; [0.1\,:\,100]~$M_\odot$; fully sampled\\
Gas covering fraction           & 0.5\\
\hline \hline
\newtag{B}{tab:params_phangs}) PHANGS--HST \\
\hline
Star formation history          & instantaneous burst\\
Reddening \& Extinction         & $E(B{-}V)=$~[0\,:\,1.5]~mag; $\Delta=0.01$~mag; $R_V=A_V/E(B{-}V)=3.1$\\
Ages                            & [1\,:\,10]~Myr with $\Delta T=1.0$~Myr; [11\,:\,13\,750]~Myr with $\Delta\log(T/{\rm Myr})=0.3$\\
SSP model                       & \cite{bruzual03} \\
Metallicity                     & $Z=0.02$ \\
IMF                             & \cite{chabrier03}; [0.1\,:\,100]~$M_\odot$; fully sampled\\
Gas covering fraction           & 0.0\\
\hline \hline
\end{tabular}
\caption{(A) Parameter ranges adopted in the benchmark comparison between \textsc{cigale} and LEGUS SED fits. The extinction curve adopted for this comparison is from \protect\cite{cardelli89} appropriate for the Milky Way. (B) Parameters adopted for PHANGS--HST SED modelling.}
\label{tab:parameters}
\end{table*}

\begin{figure}
	\includegraphics[width=\columnwidth]{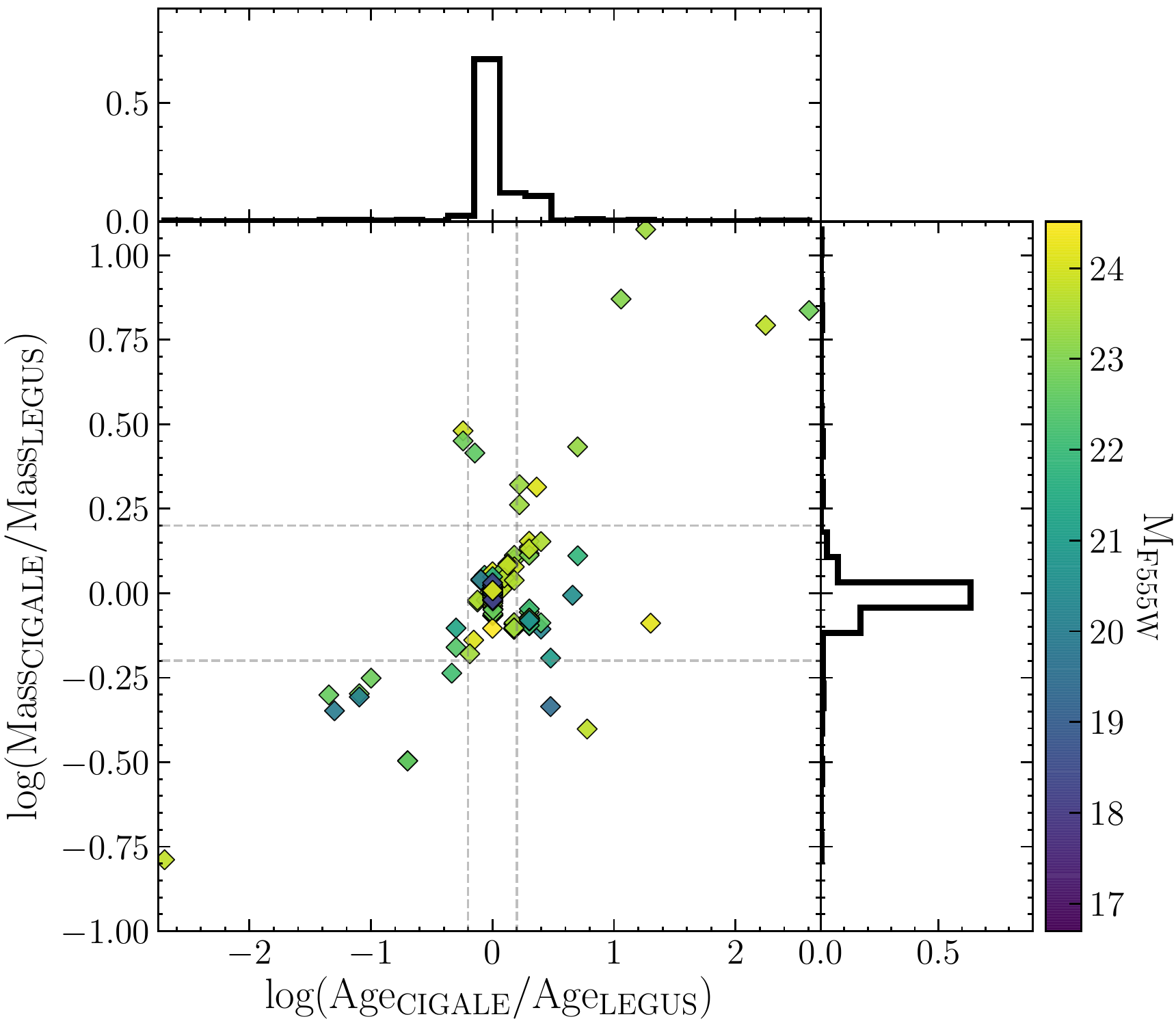}
    \caption{The residuals in \textsc{cigale} versus LEGUS stellar cluster ages and masses. Both approaches utilize the \textsc{yggdrasil} stellar tracks with a 50 per cent covering fraction (Table~\ref{tab:parameters}\ref{tab:params_legus}) and both the \textsc{cigale} and LEGUS values are based on $\chi^2$ minimization. Dotted lines mark $\pm 0.2$~dex around perfect agreement. Data are colored by $V$-band ($F555W$) Vega magnitude. The histogram above the scatter plot shows the distribution of the age ratios and the histogram to the right shows the distribution of the mass ratios. The median age ratio is $0.001 \pm 0.027$~dex. The median mass ratio is $0.003 \pm 0.011$~dex. Clusters with poor agreement are generally found to be faint with large photometric uncertainties.}
    \label{fig:residual_residual}
\end{figure}

Figure~\ref{fig:residual_residual} provides the residuals when comparing our \textsc{cigale}-derived stellar cluster ages and masses to those given by LEGUS. The agreement between the LEGUS and CIGALE SED fitting results is good. The median value of the age ratios, defined as $\log$(Age$_{\rm CIGALE}$/Age$_{\rm LEGUS}$), is $0.001 \pm 0.017$~dex. The median value of the mass ratios, defined as $\log$(Mass$_{\rm CIGALE}$/Mass$_{\rm LEGUS}$), is $0.003 \pm 0.011$~dex. Looking at the best-fitting SED models of the outliers, we do not find poor fitting models (i.e., large $\chi^{2}$ values) or inaccurate/unreliable photometry, but we find that the outliers tend to be faint as seen in Figure~\ref{fig:residual_residual}. In short, \textsc{cigale} is able to derive similar physical properties as LEGUS for the vast majority of clusters in the test case galaxy NGC~3351, establishing \textsc{cigale} as a reliable star cluster SED modeling code and consistent with previous star cluster SED fitting codes.

\section{Derivation of Physical Properties for PHANGS--HST Star Clusters}
\label{sec:physical_properties}

In this section, we review the process for deriving the masses, ages, and reddening for PHANGS--HST star clusters with \textsc{cigale}. We use the aperture and foreground Milky Way extinction corrected photometry and propagate an additional 5 per cent uncertainty within \textsc{cigale} into the photometric errors to account for systematic errors in the flux calibration.

Throughout the rest of the paper, our focus is on comparisons with the class~1 and 2 clusters (i.e., potentially bound), rather than the class~3 compact associations. Cluster classes are defined in the same way as in LEGUS -- class~1 clusters are symmetric and compact, class~2 clusters are asymmetric and compact, and class~3 clusters are multi-peaked compact associations. We note that, while we still include class~3 compact associations detected by our cluster pipeline in our catalogs, and provide visual classifications (D.\ A.\ Thilker et al. in prep. and B.\ C.\ Whitmore et al. in prep.) and SED fitting for those objects, we will use a more systematic hierarchical approach (watershed algorithm) for identification of younger star-forming associations, which is distinct from the cluster pipeline. This is detailed in K.\ L.\ Larsen et al. (in prep.) and will also include comparisons between age estimates using \textsc{cigale} for stellar associations identified with a watershed algorithm and the LEGUS and PHANGS--HST class~3 objects. The focus on class~1 and 2 clusters for this paper makes this cluster sample ``exclusive" \citep[see][for discussions on exclusive versus inclusive cluster samples]{krumholz19,adamo20}.

The set of SEDs available in the model grid is determined by the choice of input parameters. Table~\ref{tab:parameters}\ref{tab:params_phangs} summarizes our choices and the following sections discuss these choices. We study single-aged stellar populations, as the clusters do not have complex SFHs and, for the purposes of PHANGS science, can be effectively modeled by an ``instantaneous burst'' of star formation. With the functionality available in \textsc{cigale}, we implement this SFH by utilizing the double-exponential SFH module \texttt{sfh2exp} and inputting a very short e-folding time for the stellar population (a microburst) and a zero mass fraction of the second/late starburst population. This effectively acts as an instantaneous burst of star formation. We tested this method against directly fitting to the SSP models themselves and verified that the results are consistent. We also adopt solar metallicities as the PHANGS--HST galaxies are all approximately solar (Section~\ref{sec:data}). By adopting a fully-sampled IMF, \cite{chabrier03} in our case, the SED modeling is ``deterministic.'' 

As given in Section~\ref{subsec:cigale}, the model grid of ages consists of ten linearly spaced models from 1 to 10~Myr with $\Delta T = 1$~Myr and 100 log-spaced models from 11 to 13750~Myr with $\Delta\log(T/{\rm Myr}) \approx 0.3$. We check if any differences arise when assuming a fully linearly sampled age grid and find no significant changes to the resulting cluster properties even for the bimodal PDF cases.

\subsection{Single-Age Stellar Population Models \& Nebular Emission}
\label{sec:ssp}

\begin{figure}
	\includegraphics[width=\columnwidth]{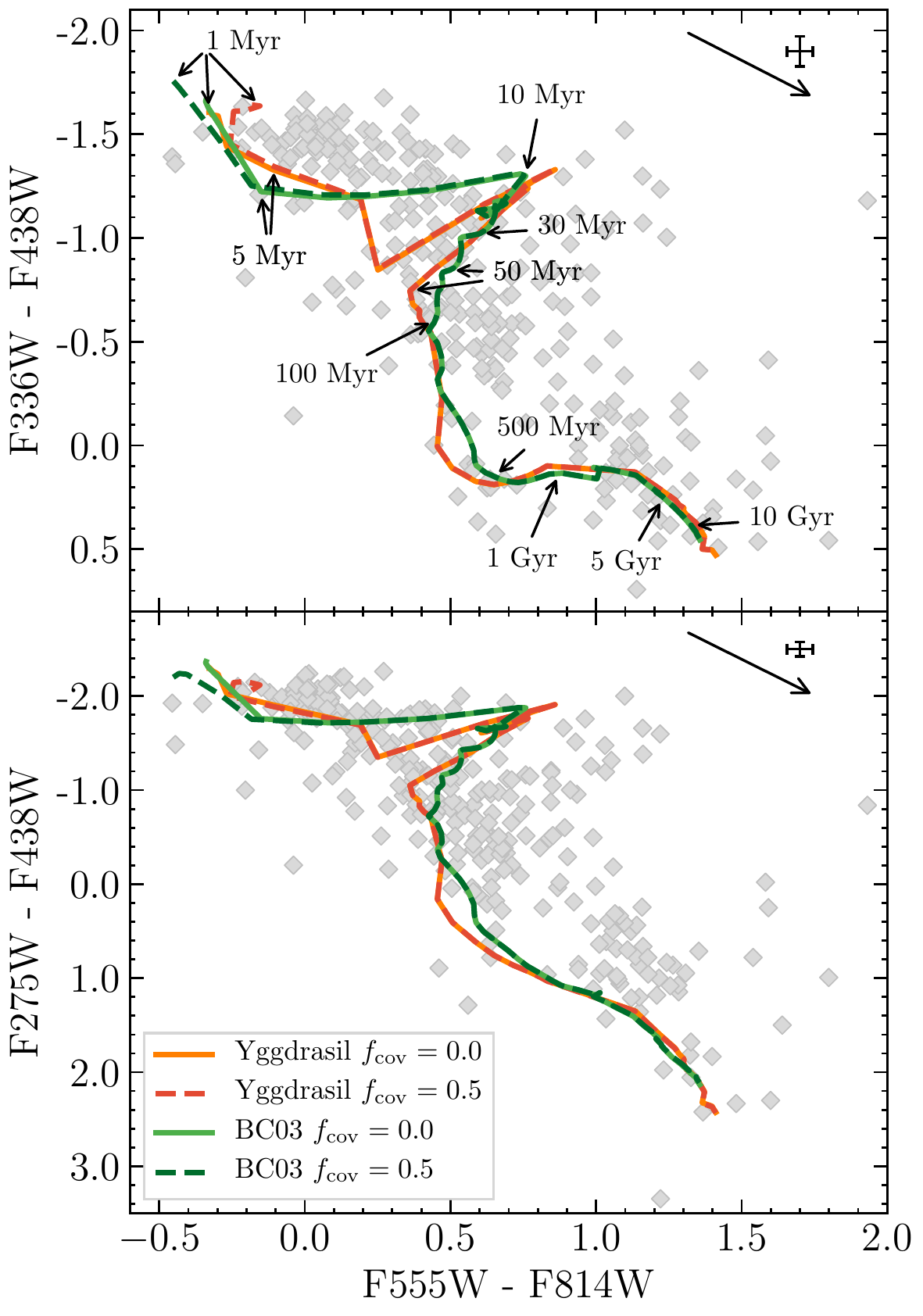}
    \caption{Top: $F336W{-}F438W$ vs $F555W{-}F814W$ ($U{-}B$ vs $V{-}I$) color-color diagram for a collection of theoretical single-age stellar population models, along with the PHANGS--HST stellar clusters for NGC~3351. Magnitudes are in the Vega system. The stellar clusters have been aperture corrected as well as corrected for foreground Milky Way extinction. The \citetalias{bruzual03} (green) and \textsc{yggdrasil} (orange) tracks are shown both with (dashed) and without (solid) nebular emission. The y-axis has been flipped so the top left-hand corner of the diagram is the `bluest' and the bottom right-hand corner is the `reddest'. The top panel annotates various ages along the SSP tracks; ages where the tracks deviate significantly from each other are pointed out with two arrows. The class~1 and 2 star clusters are plotted in grey. Both panels include a reddening vector for $A_V=1$~mag as well as a typical error bar for the cluster photometry computed from the median uncertainty in the colors. Bottom: Same as the top panel except with $F275W$ ($NUV$) instead of $F336W$ ($U$).}
    \label{fig:color_color_tracks}
\end{figure}

We plan to consistently use the same single-aged stellar population models as the baseline for PHANGS--HST stellar cluster analyses. In this section, we compare two common models: \citetalias{bruzual03} and \textsc{yggdrasil} \citep{zackrisson11}, the latter of which were adopted for the SED fitting of the LEGUS star clusters \citep{adamo17}. Tracks from both models are shown in Figure~\ref{fig:color_color_tracks} overplotted on the PHANGS--HST class~1 and 2 cluster photometry. Included in the figure are the model tracks with and without nebular emission, specified by a parameter describing the gas covering fraction. The gas covering fraction, opposite of the escape fraction, determines the fraction of the Lyman continuum photons that ionize the surrounding gas. Nebular continuum and line emission can significantly contribute to the observed fluxes within broadband filters, especially for spatially integrated observations of star-forming galaxies \citep[e.g.,][]{anders03,groves08,boquien10,reines10,pellegrini20b}. PHANGS--HST observations, however, benefit from the combination of \textit{HST}'s angular resolution and relative proximity of PHANGS galaxies, such that star clusters and surrounding \HII\ regions can often be spatially disentangled, except for (1) crowded regions where it is still often unclear which star clusters are responsible for ionizing a given patch of H$\alpha$ emission, and (2) clusters with compact H$\alpha$ morphologies \citep{hannon19}. The \textsc{yggdrasil} tracks with zero covering fraction (i.e., no nebular emission) are relatively consistent with the \citetalias{bruzual03} tracks with no nebular emission; the main difference is in between 5 and 10~Myr where the \textsc{yggdrasil} tracks dip to much redder colors than the \citetalias{bruzual03} tracks. The addition of nebular emission is insignificant for most ages in both \textsc{yggdrasil} and \citetalias{bruzual03}. As expected, only for the youngest ages ($\sim$0$-5$~Myr) is there a difference: a hook-like feature towards redder colors in the \textsc{yggdrasil} track with a covering fraction of a half while the \citetalias{bruzual03} track with nebular emission branches towards slightly bluer colors. 

While both the \textsc{yggdrasil} and \citetalias{bruzual03} tracks show good agreement with the photometry in color-color space as is apparent in Figure~\ref{fig:color_color_tracks}, we decide to adopt the \citetalias{bruzual03} tracks because they were originally designed to match galactic star clusters as well as reproduce the colors of star clusters in the Magellanic Clouds. The \textsc{yggdrasil} stellar population models were developed for constraining high redshift galaxies \citep{zackrisson11}.\footnote{We note that both the \citetalias{bruzual03} and \textsc{yggdrasil} models do not account for possible binary stars within the clusters.} 

\begin{figure*}
	\includegraphics[width=\textwidth]{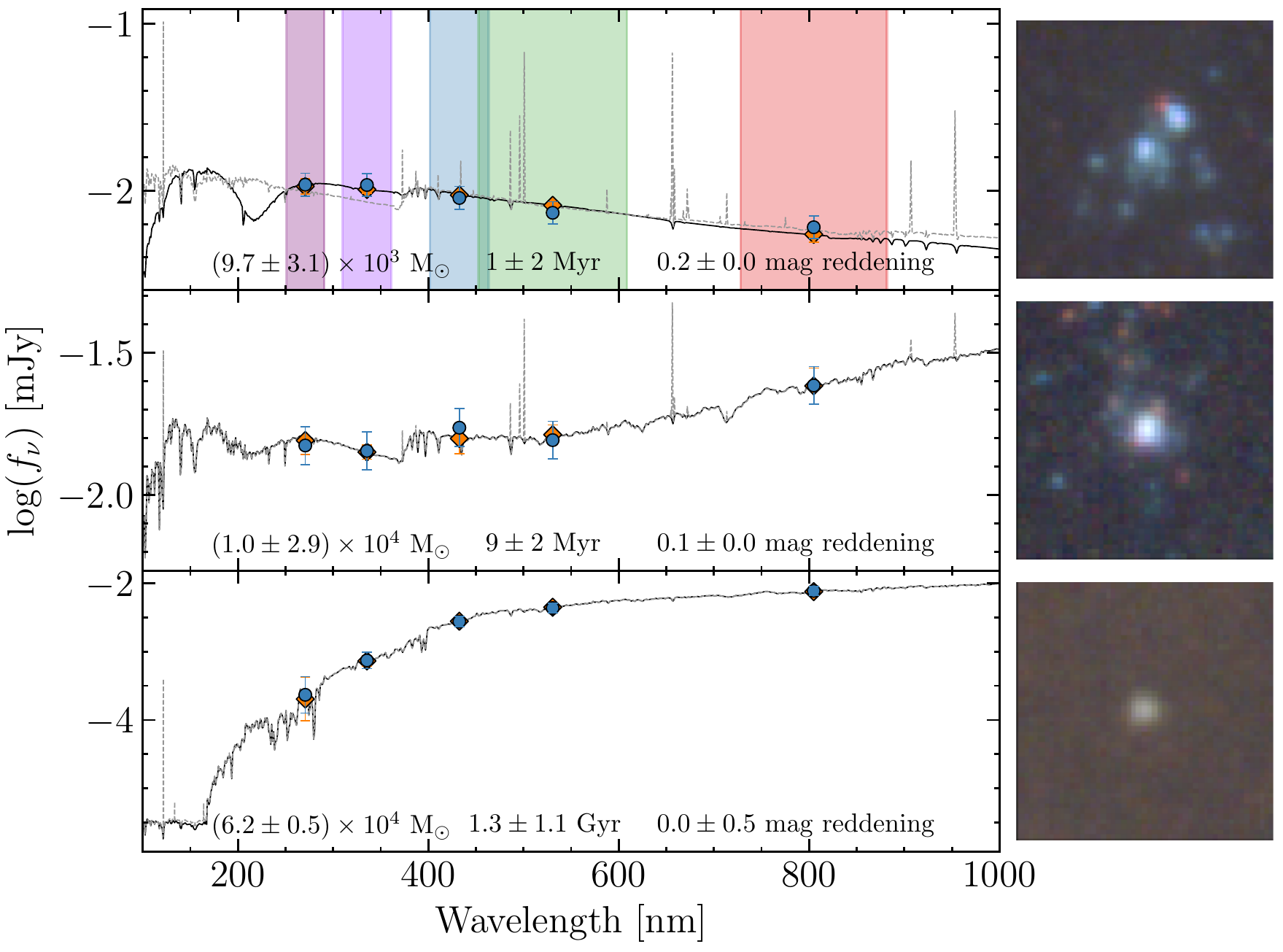}
    \caption{Example \textsc{cigale} fits (black line) for three PHANGS clusters (two young, one old) in NGC~3351. Blue circles mark the observed fluxes and orange diamonds indicate the fluxes extracted from the best-fit model. Error bars denote $3\sigma$ uncertainties on the observation and model fluxes. In some cases, the error bars are smaller than the data markers. Colored strips in the top panel illustrate the filter bandpass rectangular widths. The black solid line is the best-fitting SED model using the parameters outlined in Table~\ref{tab:parameters}\ref{tab:params_phangs}; the grey dashed line is the same SED model while including the nebular emission component ($f_{\rm cov} = 0.5$). Best-fitting star cluster masses, ages, and reddenings are given for each cluster. A $1\arcsec\!\!\times\!1\arcsec$ false-color ($B$-$V$-$I$) image of the cluster is given to the right of each SED. The same image stretch is used for all three images to allow for easy visual comparison.}
    \label{fig:sed_fit_example}
\end{figure*}

We check how the inclusion of a nebular emission component into the SED fitting affects the results. In Figure~\ref{fig:color_color_tracks}, we see the \citetalias{bruzual03} model tracks only differ slightly at very young ages when including the nebular emission. Figure~\ref{fig:sed_fit_example} shows the best-fitting SED models for three example PHANGS--HST clusters. The grey dashed line is the best-fitting SED model while including a nebular emission component. In this case, we assume an ionization parameter $\log{U}=-2.0$, line width of 300 km s$^{-1}$, and a covering fraction $f_{\rm cov} = 0.5$ (i.e., 50 per cent of the Lyman continuum photons ionize the surrounding gas). The first two clusters in Figure~\ref{fig:sed_fit_example} are young, blue clusters where the nebular emission could have the largest impact on the SED fitting. The first cluster's SED fit with nebular emission gives a larger reduced $\chi^{2}$ value (8.2) than compared to the non-nebular fit (2.44). For the second cluster, we find similar reduced $\chi^{2}$ values between the nebular and non-nebular SED fits. As a contrast to these young clusters, the best-fitting SED for an old, red cluster is also given in the bottom panel of Figure~\ref{fig:sed_fit_example}. The difference between SEDs with and without nebular emission is negligible. Therefore, for the PHANGS--HST sample, we use SED fitting without a nebular emission component. We note that \cite{krumholz15} perform a similar analysis and find the resulting cluster properties are robust to choice of evolutionary track and the inclusion of a nebular emission.

\subsection{Dust}
\label{sec:dust}

For the PHANGS--HST model grids, we make use of a Milky Way-like extinction curve from \cite{cardelli89} with $R_{V} = 3.1$. \cite{krumholz15} find that Milky Way-like extinction curves provide better results over other models for solar metallicity, face-on spiral galaxies when working with deterministic models. We allow the internal reddening to range from $E(B{-}V) = 0$ to $1.5$ mags in 0.01~mag steps -- the same range adopted by LEGUS.\footnote{The cluster photometry is corrected for foreground Milky Way extinction before the SED fitting.} For reference, the drift scan spectroscopy for NGC~3351 from \cite{moustakas10} indicates $E(B{-}V)$ for nuclear (aperture of $\sim$6~arcsec$^2$), circumnuclear (400~arcsec$^2$), and large-scale radial strips ($\sim$few arcmin$^2$) of $0.03\pm0.05$~mag, $0.64\pm0.02$~mag and $0.55\pm0.07$~mag, respectively. Though the regions sampled by \cite{moustakas10} exhibit smaller values of reddening than $E(B{-}V)_{\rm max}=1.5$~mag, we have chosen to keep the maximum value since sight lines to some individual clusters will inevitably exceed the large-scale spatial averages. As a sanity check, we perform a test SED fitting run with a maximum reddening of 3.0~mag allowed and find only two clusters with a reddening greater than 1.5~mag (1.72 and 1.74~mag). Therefore, we decide to use a maximum value of $E(B{-}V)=1.5$~mag over smaller values to allow for individual sight lines with large reddening. 

\begin{figure}
    \centering
    \includegraphics[width=\columnwidth]{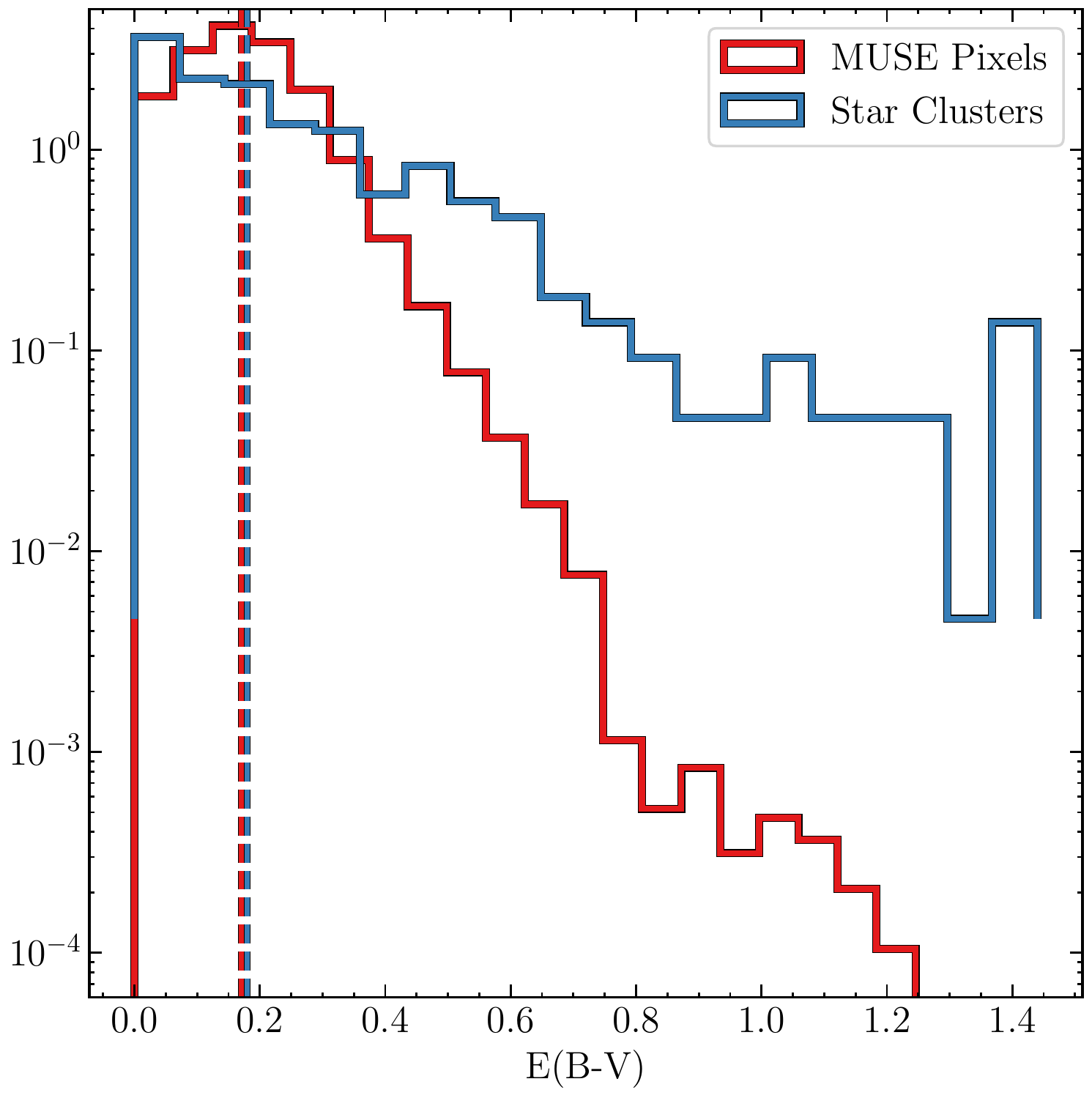}
    \caption{Normalized logarithmic distributions of $E(B{-}V)$ values for the pixels (0.2$\arcsec$ in size) across the PHANGS--MUSE map of NGC~3351 as determined by Balmer decrement measurements (red) and the individual star cluster best-fitting reddenings as found by our SED fitting (blue). Only the reddening values of star clusters found where the Balmer decrement has been measured are used. Colored dashed lines mark the median values for each distribution: 0.171~mag for the MUSE pixels and 0.18~mag for the star clusters.}
    \label{fig:musebv}
\end{figure}

It could be possible to derive accurate maximum reddening values for the youngest ($\lesssim$10~Myr) stellar populations using the PHANGS--MUSE integral field unit (IFU) data and Balmer decrement measurements. The IFU maps cover an area of the galaxies similar to our \textit{HST} footprint but, the ground-based MUSE $\sim$1\arcsec\ angular resolution is too coarse to be directly applicable to our HST-resolution maps ($\sim$0.08\arcsec). Still, the $A_V$ maps generated from Balmer decrement measurements from PHANGS--MUSE (E.\ Emsellem et al. in prep.) can be used as a independent check of our allowable reddening values. $E(B{-}V)$ values for each pixel in the PHANGS--MUSE map are computed by assuming $R_V=A_V/E(B{-}V)=4.05$ \citep{calzetti00}. The distribution of the reddening values by pixel is given in Figure~\ref{fig:musebv} which shows a roughly power-law distribution with a median value of 0.171~mag and a maximum value of 1.24~mag. This distribution aligns fairly well with the chosen reddening parameters for the SED fitting. 

A well known problem in all star cluster SED fitting work is how to deal with dust extinction to break the age--reddening degeneracy. A star cluster can appear to be red due to age or due to reddening by dust. In our sample, we find $\sim$70 (20 per cent) of the clusters appear to be bimodal in the age and reddening PDFs. There are a number of possible ways to move towards breaking the degeneracy. In Section~\ref{sec:prior}, we make use of Bayesian priors as an attempt to help limit the number of degenerate cases. See Section~\ref{sec:future} for a discussion on additional methods we might employ in the future to break the age--reddening degeneracy in the PHANGS--HST sample.

\subsection{Use of Fluxes vs. Magnitudes in SED Fitting}

SED fitting can be carried out inputting either linear flux units \citep[e.g.,][]{dacunha08} or logarithmic magnitude units \citep[e.g.,][]{chandar10b}. We explore here if significant differences arise in key extracted output parameters (cluster age, mass, and reddening) when utilizing fluxes versus magnitudes as input to the SED fits. To perform the SED fitting in logarithimic magnitude units, we use the \textsc{cigale} branch \texttt{SSPmag} which expects the input photometry to be of the form $2.5\log(f_{\nu})$. The uncertainties must be properly converted from linear fluxes. Comparing the best-fitting results from fluxes to those from magnitudes reveals they are quantitatively very similar. The median age residual, defined as $\log({\rm Age_{\rm flux}} / {\rm Age_{\rm mag}})$, is $0.000 \pm 0.026$~dex. The median mass residual is $-0.001 \pm 0.011$~dex, and the median reddening residual is $0.000 \pm 0.009$~mag. The main difference between the two results is that the magnitude-based fits have, on average, much larger uncertainties for the ages, masses, and reddenings.

Understanding how the uncertainties in our flux measurements are distributed---preferentially Gaussian in linear or logarithmic units---is essential for determining if fitting in flux or magnitudes is more accurate. However, we find the distribution of the uncertainties to be similarly non-Gaussian for both fluxes and magnitudes for all five filters; all distributions display similar positive skewness with long tails out to high uncertainties. We cannot come to a conclusion on using fluxes or magnitudes based on their uncertainty distributions. A more thorough study of the photometric uncertainties is presented in D.\ A.\ Thilker et al. (in prep.). An advantage of working in linear fluxes is in the case of non-detections (i.e., insignificant positive flux measurements or even slightly negative measurements). In linear fluxes, meaningful information can still be fed into the SED fitting using flux upper limits which is not as easily accomplished with magnitudes. Additionally, at low signal-to-noise, errors in magnitude units are not expected to be symmetric and the $\chi^{2}$ minimization assumes errors to be symmetric and Gaussian \citep{hogg10}. Hence, these assumptions will not hold when using logarithmic units in the low signal-to-noise regime. Given these results, we choose to perform all the SED fitting in linear fluxes.

\subsection{Summary of SED Modelling Assumptions}
We summarize our SED modelling assumptions here and in Table~\ref{tab:parameters}\ref{tab:params_phangs}. The age model grid consists of ten linearly-spaced models from 1~Myr to 10~Myr with $\Delta T=$~1~Myr and 100 log-spaced models from 11~Myr to 13\,750~Myr with $\Delta\log(T/{\rm Myr})\approx0.3$. We assume a Milky-Way like extinction curve \citep{cardelli89} with $R_V = 3.1$. The reddening $E(B{-}V)$ model grid spans from 0~mag reddening to 1.5~mag with $\Delta E(B{-}V) = 0.01$~mag. This gives a final age-reddening model grid with 16\,610 models. Each model on the grid has a corresponding mass based on the \citetalias{bruzual03} SSP track and the fully-sampled \citet{chabrier03} IMF. We assume solar metallicity and a gas covering fraction of zero.

\section{Results \& Discussion}
\label{sec:results}

\begin{figure}
    \centering
    \includegraphics[width=\columnwidth]{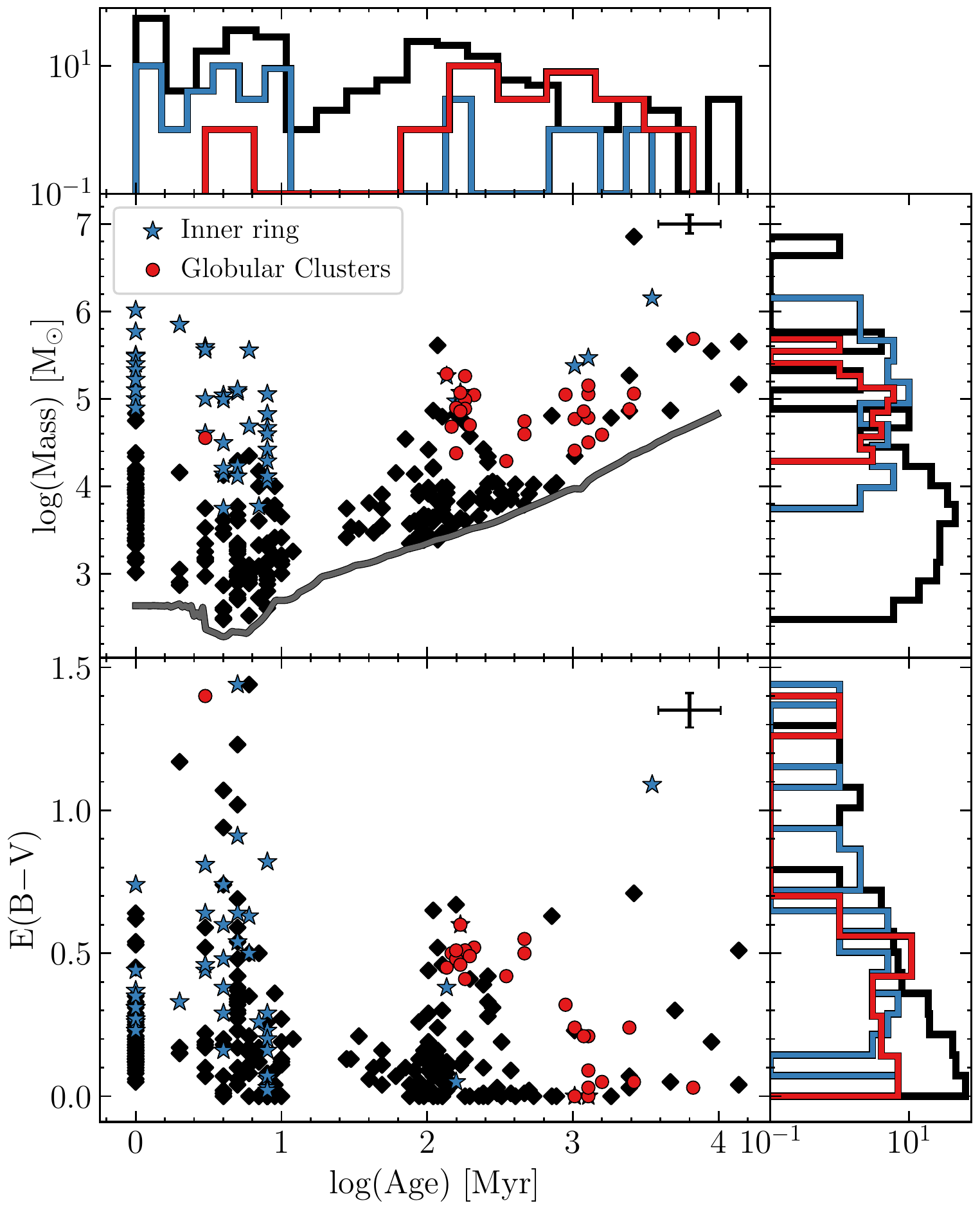}
    \caption{Best-fitting cluster masses and reddenings versus ages for the visually-classified class~1 and 2 clusters of NGC~3351. Clusters of the inner star-forming ring are shown as blue stars and visually identified globular cluster candidates are shown as red circles. Logarithmic histograms beside each axis give the distributions of the cluster properties. Typical error bars, computed from the median of property uncertainties, are given in the corner of the two panels. The inner ring star clusters are found to be mostly young and massive with a large range in reddening. All but one of the globular cluster candidates are found to have ages $>100$~Myr and to be massive following along the $M_{V}=-6$ Vega mag observational limit (grey line).}
    \label{fig:agemass}
\end{figure}

We provide a first presentation of the star cluster ages, masses, and reddenings based on the $\chi^{2}$ minimized best-fitting results for NGC~3351 using PHANGS--HST photometry and the \textsc{cigale} SED fitting as described in Section~\ref{sec:physical_properties}. Figure~\ref{fig:agemass} shows the derived cluster masses and reddening values as a function of the cluster ages as well as the distributions of the cluster properties. We find a large population of young clusters at 1~Myr with reddening values ranging from 0.05~mag to 0.74~mag. The most highly reddened clusters are within ages from $\sim$3~Myr to 10~Myr. A noticeable dearth of clusters at ages from 10 and 30~Myr is found due to the inherent degeneracy of the SSP model at these ages where clusters of similar colors can be found to be either above or below 10~Myr with corresponding reddening. The $\chi^{2}$ minimization seems to prefer the younger models with slightly higher extinctions. We also find a trend of the older clusters to be more massive, an expected selection effect, due to the absolute magnitude limit of the observations of around $M_{V} = -6$~mag. As star clusters grow older, they will become less luminous due to the evolution of the stellar population.

\begin{figure}
    \centering
    \includegraphics[width=\columnwidth]{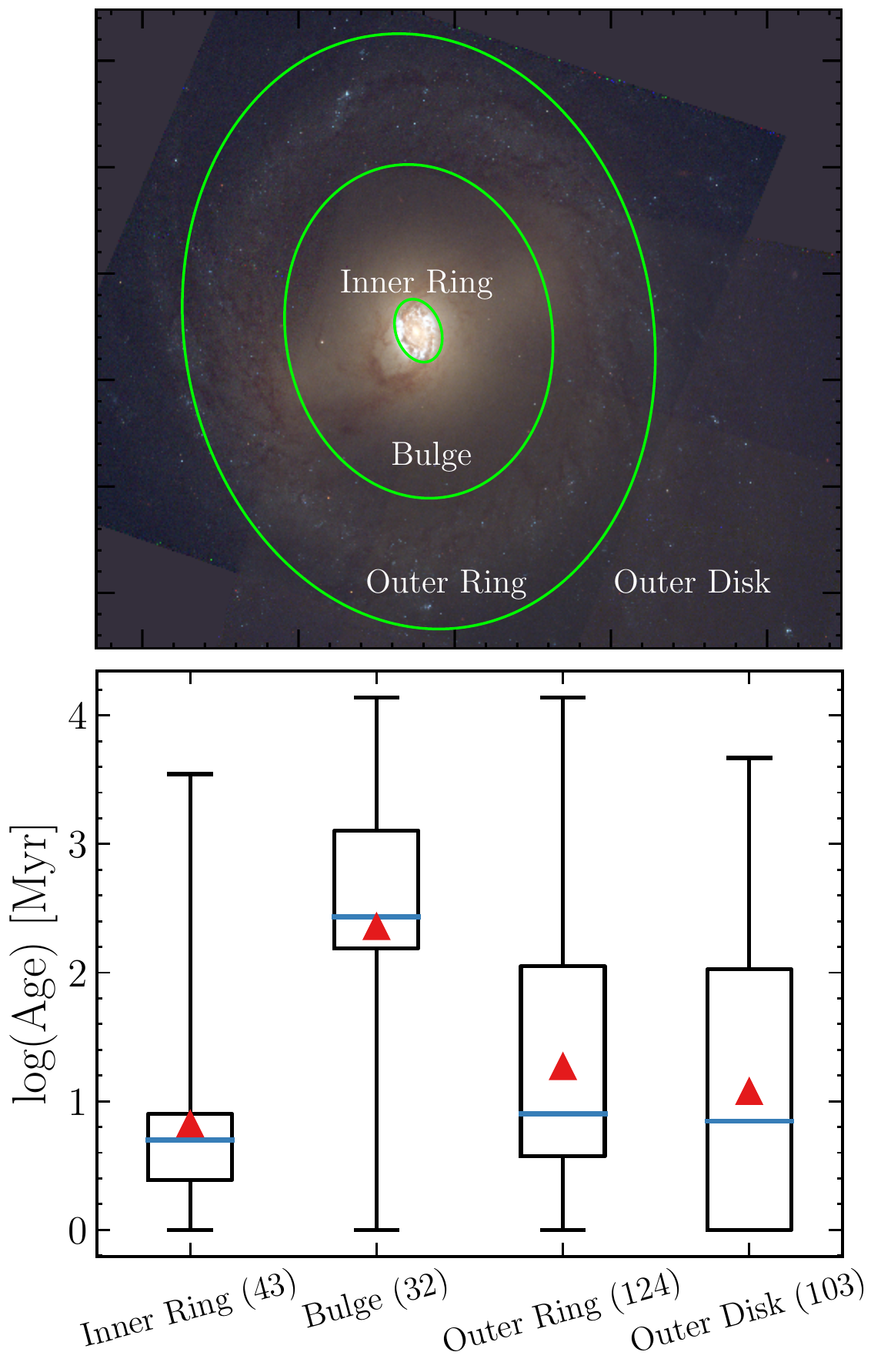}
    \caption{Top: Color image of NGC~3351 with the four regions outlined and labeled. Bottom: Box and whisker plot showing the distribution of the star cluster ages within the four regions moving radially outward from the center: the inner star-forming ring (43 clusters), the inter-ring or bulge region (32 clusters), the outer star-forming ring (124 clusters), and the outer disk beyond the outer ring (103 clusters). The blue lines in each box mark the median best-fitting cluster age and the red triangles mark the mean. Boxes extend from the lower to upper quartiles and the whiskers extend to the minimum and maximum ages found in each region.}
    \label{fig:box}
\end{figure}

\begin{figure*}
    \centering
    \includegraphics[width=\textwidth]{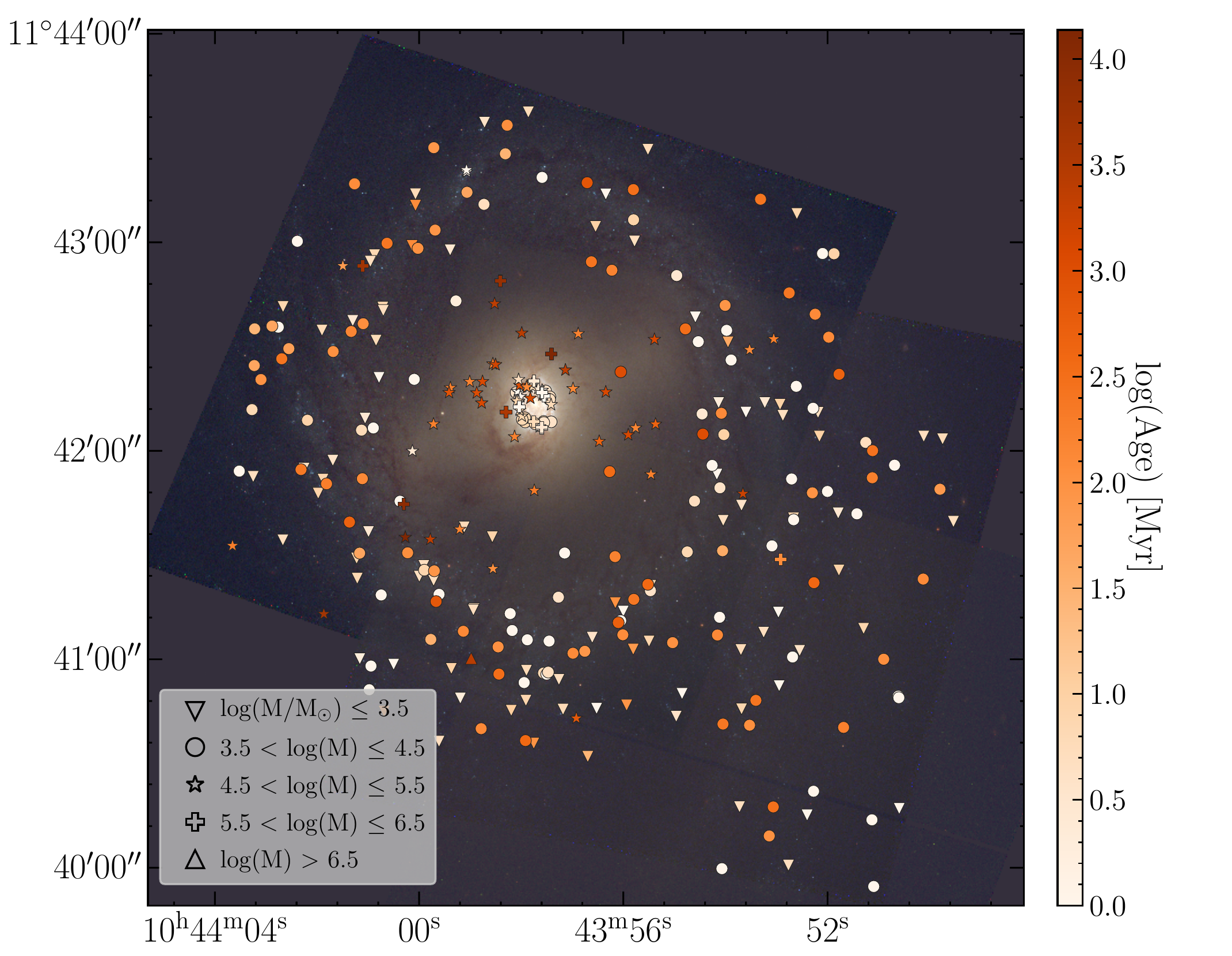}
    \caption{Color image of NGC~3351 with the location of the clusters overplotted. Clusters are color-coded by best-fitting log(Age/Myr). Data point symbols differentiate between five log(Mass/$M_{\odot}$) bins. The majority of clusters found in the outer ring and disk fall in the two lowest mass bins (downward triangles and circles). Higher mass clusters (stars and crosses) are generally found to be in the stellar bulge and the inner star-forming ring. The most massive cluster (upward triangle) is located in the outer ring.}
    \label{fig:map_cluster_props}
\end{figure*}

Given the morphology of NGC~3351, we can break down the galaxy into distinct regions, allowing us to probe different star-forming environments within a single galaxy. Environmental masks have been developed by the PHANGS collaboration (M.\ Querejeta et al. in prep.) identifying disks and bulges using the Spitzer Survey of the Stellar Structure in Galaxies (S$^4$G) pipeline as well as rings, bars, and lenses from \cite{herrera-endoqui15}. The box and whisker plot in Figure~\ref{fig:box} shows the breakdown of the star cluster ages in four distinct regions, moving radially outward from the center of the galaxy: the inner star-forming ring with 43 clusters, the inter-ring or stellar bulge region with 32 clusters, the outer star-forming ring with 124 clusters, and the outer disk region beyond the outer ring with 103 clusters. Median best-fitting cluster log(Age/Myr) for the four regions are 0.69, 2.43, 0.90, and 0.84 and the means are 0.82, 2.36, 1.27, and 1.08. We find all four regions to have a similar dynamic range of ages but the bulge is home to significantly older star clusters on average. Over time, star clusters tend to migrate away from their natal gas clouds and disperse more uniformly with age over the galaxy \citep{gielesBastian08,bastian09,davidge11,kruijssen11,grasha19}. The bulge has few signatures of current star formation, hence any star clusters found in that area must be old enough to have had the time to migrate there or have been left in place by past star formation.

We show the spatial distribution of the clusters and their properties across the galaxy in Figure~\ref{fig:map_cluster_props}. The majority of the clusters found in the outer ring and disk have masses less than $10^{4.5}$~$M_{\odot}$. Higher mass clusters are generally found to be in the stellar bulge and inner star-forming ring. The majority of clusters in the inner ring are young as well. \cite{kruijssen14} and \cite{reina-campos17} predict higher mass clusters at small galactocentric radii due to higher ISM pressure which our result supports. These young, massive clusters can also be formed from the funneling of gas into the center by the galactic bar \citep{swartz06}. Approximately 70 per cent of the clusters have reddenings less than 0.3~mag ($A_{V} < 0.93$~mag). There is most likely a population of faint, low mass clusters in the inner ring which we are missing due to the bright background level. For the more heavily reddened and dust-obscured clusters, a high-resolution study with \textit{James Webb Space Telescope} would provide a more robust examination of the dust. We note that the most massive cluster is found in the outer ring. 

There are valid concerns regarding the ages of intermediate and older ($>$100~Myr) clusters due to the age-reddening degeneracy. In Section~\ref{subsec:glob}, we will discuss the how the SED fitting handles clusters which are likely old globular clusters. Here, we check on the intermediate-aged (100~Myr to 1~Gyr) clusters based on their location in color-color space (Figure~\ref{fig:color_color_tracks}) and location within the galaxy. A cut in color-color space is made to isolate the clusters redder than (i.e., below) the \citetalias{bruzual03} track at $\sim$50~Myr. This gives 135 clusters with 25 clusters ($\sim20\%$) found to have ages of 10~Myr and younger and corresponding high reddening values. Nine of these clusters are found within the dusty inner ring of the galaxy meaning they are likely highly reddened as the SED fitting results suggest. One cluster is found in the bulge and the remaining 15 are found in the outer ring and disk. In these regions, it is unlikely to have the high reddening as returned by the SED fitting. Overall, this represents only $5\%$ of the clusters (16 out of the total 302) which are likely intermediately aged (based on the location in the color-color diagram) but given incorrect young ($<$10~Myr) ages. See Section~\ref{sec:future} for a discussion on how we may resolve this issue.

\subsection{Comparison with LEGUS}

There are 199 visually-classified class~1 and 2 PHANGS--HST clusters that are also found in the LEGUS cluster catalog. On average, our best-fitting ages agree quite well with a median logarithmic age ratio of $0.000 \pm 0.067$~dex. We find our cluster masses to be slightly larger compared to LEGUS with a median logarithmic mass ratio of $0.035 \pm 0.035$~dex and our cluster reddenings following the same trend with a median $E(B{-}V)$ difference of $0.030 \pm 0.020$~mag. There is no correlation of the residuals with ages, masses, or reddenings.

We can also examine agreement in the four different regions as discussed in the previous section. We find better agreement of the best-fitting cluster ages in the outer ring (median residual = $0.050 \pm 0.125$~dex; $N=70$~clusters) and outer disk ($0.0 \pm 0.077$~dex; $N=67$) than compared to the inner ring ($-0.125 \pm 0.131$~dex; $N=33$) and bulge ($0.313 \pm 0.217$~dex; $N=29$). For the bulge clusters, the best-fitting reddening is slightly lower compared to LEGUS with an median residual of $-0.040 \pm 0.070$~mag, while the inner ring ($0.070 \pm 0.039$~mag), the outer ring ($0.005 \pm 0.037$~mag), and outer disk ($0.060 \pm 0.021$~mag) show good agreement if not slightly higher. The masses are well matched with LEGUS across all regions of the galaxy. 

We note that in a recent study of LEGUS star clusters in NGC~4449, \cite{whitmore20} show issues with LEGUS cluster ages and reddenings. In particular, LEGUS finds young ages for spectroscopically confirmed old globular clusters. This issue should be considered while comparing our cluster results with those from LEGUS. 

\subsection{Clusters with HII regions}

\textit{HST} H$\alpha$ imaging was obtained by H$\alpha$--LEGUS for the LEGUS pointing of NGC~3351 (PI: R.\ Chandar; S.\ Hannon et al. 2020, in prep.). We adopt the following classification scheme for the H$\alpha$ emission for each star cluster: (i) H$\alpha$ emission directly on top of cluster, (ii) H$\alpha$ emission in a ring around the cluster, (iii) H$\alpha$ emission possibly associated with the cluster, or (iv) no H$\alpha$ detected. We check our ability to accurately recover the ages of these young clusters by looking at the clusters which have strong H$\alpha$ emission directly on top of the cluster or in a ring around the cluster. There are 60 class~1 and 2 clusters with such H$\alpha$ emission which implies that they must be young (< 10~Myr). The PHANGS--HST SED modelling recovers best-fitting ages of 10~Myr and younger for 93 per cent of these clusters which bodes well for the validity of our results for young clusters. 

\subsection{Globular Clusters}
\label{subsec:glob}

\begin{figure}
    \centering
    \includegraphics[width=\columnwidth]{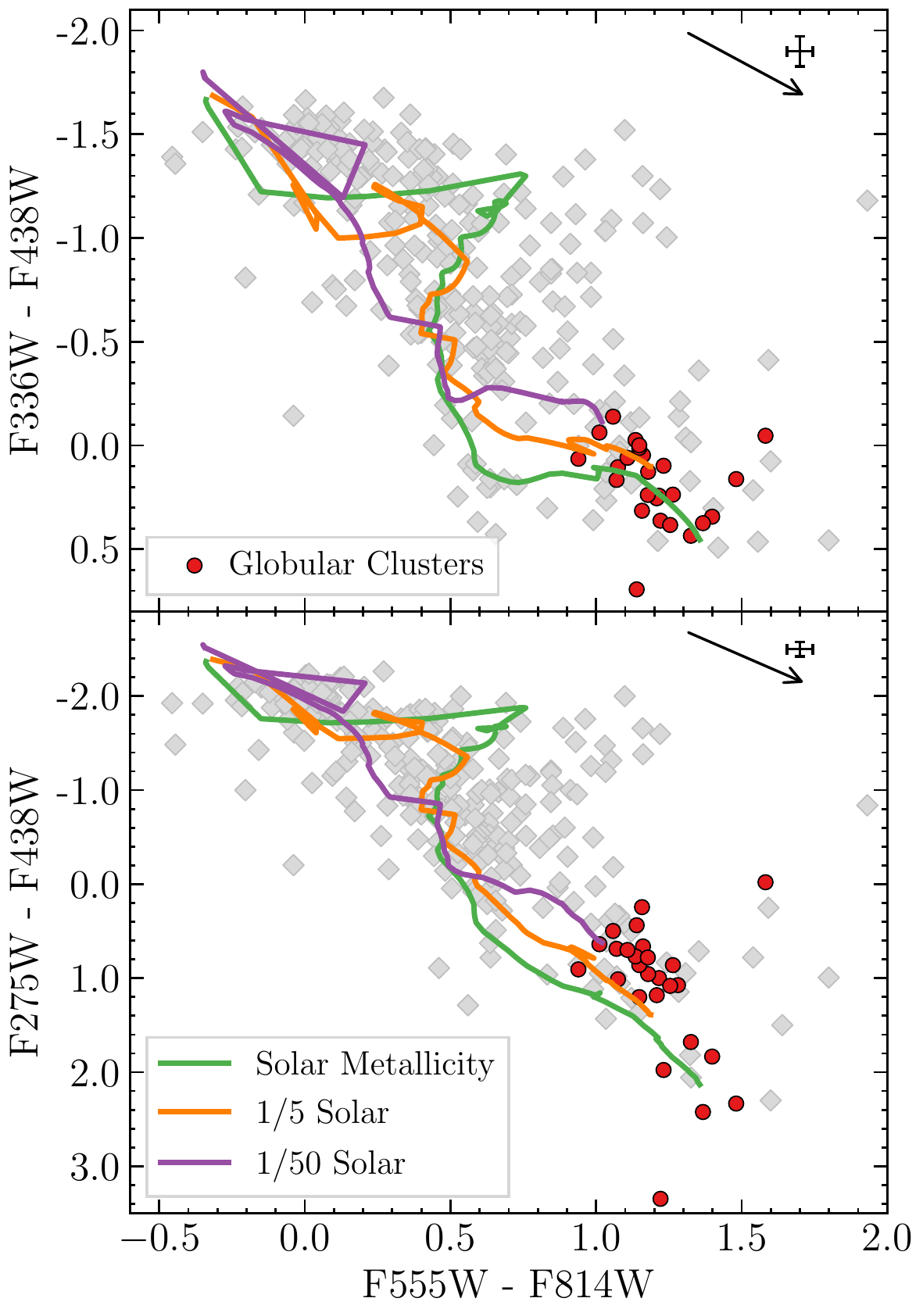}
    \caption{Color-Color diagram similar to Figure~\ref{fig:color_color_tracks} highlighting the visually-identified globular clusters candidates (red circles). The green track is the \citetalias{bruzual03} SSP model with solar metallicity used in the SED fitting. The orange track is the \citetalias{bruzual03} SSP model with $1/5$ solar metallicity and the purple track is $1/50$ solar metallicity. Reddening vectors are given in each panel along with typical error bars on the cluster colors computed by the median uncertainties of the photometry. As expected, globular cluster candidates are found at the oldest end of the SSP track. However, the sub-solar metallicity tracks provide a better match to the globular cluster candidates' colors. This is especially apparent in the bottom panel where the grouping of globular cluster candidates occupy the region at the very end of the $1/50$ solar metallicity track.}
    \label{fig:cc_glob}
\end{figure}

We have visually identified a sub-sample of candidate globular clusters within our star cluster sample for NGC~3351 in order to check if the results provided by our SED fitting are as expected for globular clusters, i.e., old ($\sim$10~Gyr) with little reddening. This is done by selecting clearly resolved objects with similar red colors in the bulge of the galaxy (see Figure~\ref{fig:box}). We note that nearly all the objects in this region are red in color with almost no blue (i.e., young) objects in the vicinity.

Figure~\ref{fig:cc_glob} shows the location of the globular cluster candidates in the \textit{UBVI} color-color diagram. The globular cluster candidates all group together in a tight region near the old end of the SSP track which demonstrates that they are indeed likely to be globular clusters. We note that the SED fitting derived ages for most of the candidate globular clusters are younger than the expected $\sim$10~Gyr (Figure~\ref{fig:agemass}). This is a well-known problem with most SED fitting approaches to age-dating cluster populations when the focus is on the young population. It is caused by two effects: (1) using high metallicity model isochrones appropriate for young, but not old, populations; (2) the age-reddening degeneracy which allows the SED fitting to find a better fit with large reddening values even in cases where it is not physically reasonable (e.g., there is very little dust in the bulge region used to select the candidate globular clusters). These two effects can compound each other since the use of the wrong isochrones for low-metallicity objects results in a gap between the main grouping of points in the color-color diagram (most clearly seen in the bottom panel of Figure~\ref{fig:cc_glob}) and where the solar-metallicity isochrone would predict a 10~Gyr cluster should be. This increases the number of cases where a better fit is found using a large reddening value (i.e., backtracking down the reddening vector) to an age of around 100~Myr on the isochrone (see Figure~\ref{fig:color_color_tracks}). These issues are discussed in more detail in \citet{whitmore20} where the same phenomena of understanding ages for old globular clusters is found for sample of spectroscopically age-dated globular clusters in NGC~4449.

Figure~\ref{fig:cc_glob} also shows sub-solar metallicity \citetalias{bruzual03} SSP tracks. We find roughly half of the globular cluster candidates benefit (i.e., return ages $\sim$10~Gyr) from the use of a lower metallicity model. The other half of the candidates must be suffering from the age-reddening degeneracy. We test this by enforcing the reddening $E(B-V)$ to zero \citep[following][]{whitmore20} and the globular cluster candidates with ages originally around 100~Myr are shifted 1~Gyr and older. Given these results, we must be cautious using the SED derived ages of globular clusters due to both the metallicity and age-reddening degeneracy effects.

\subsection{Mass Functions}
\label{subsec:mf}

\begin{figure*}
	\includegraphics[width=\textwidth]{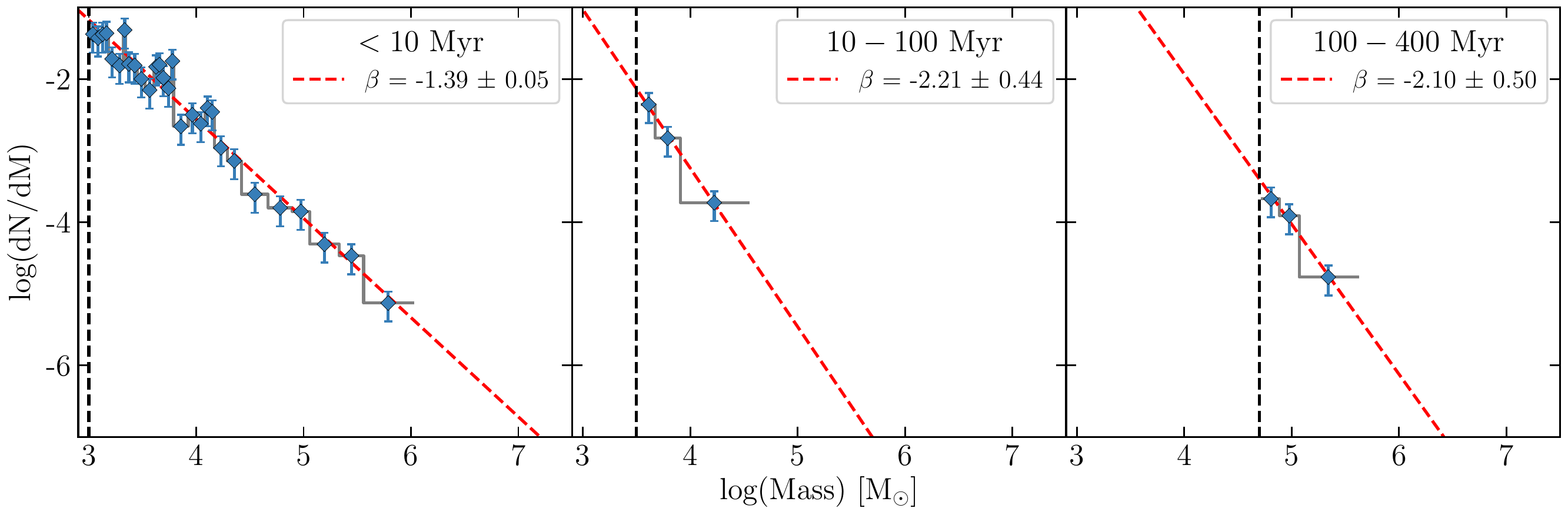}
    \caption{Cumulative stellar cluster mass functions for three age bins: less than 10~Myr (left column), 10--100~Myr (middle column), and 100--400~Myr (right column). The red dashed lines are the best-fitting power laws for each mass function with the power law exponent $\beta$ given in the top right of each panel. The black vertical dashed lines mark the completeness limits.}
    \label{fig:mass_function}
\end{figure*}

\begin{figure*}
	\includegraphics[width=\textwidth]{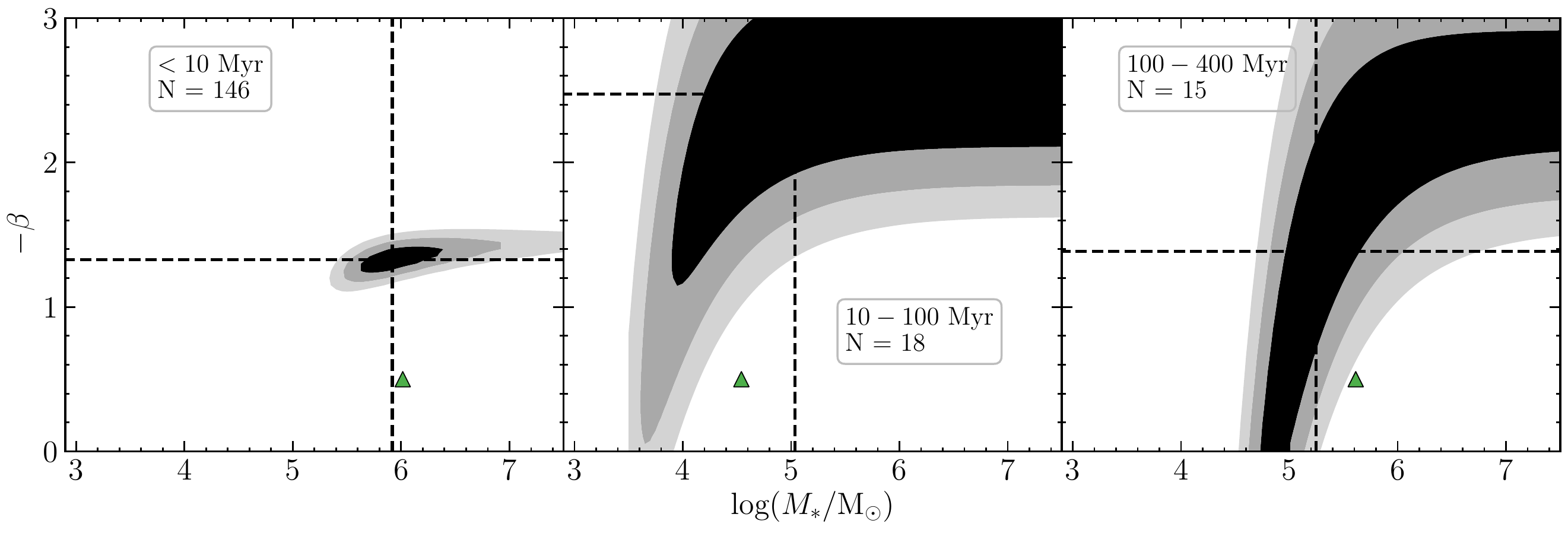}
    \caption{The 1, 2, and 3$\sigma$ confidence contours from the maximum likelihood fits for the mass functions in Figure~\ref{fig:mass_function}. The green triangles denote the position of the most massive cluster in each respective sample.}
    \label{fig:ML}
\end{figure*}

The shape of the star cluster mass function provides important information on the cluster formation and evolution. These distributions can be described to first order by a power law, $dN/dM \propto M^{\beta}$, where $\beta\approx-2$ has been found for the young cluster populations in a number of galaxies \citep[e.g.,][]{lada03,portegies10,krumholz19}. There have also been claims for an exponential-like downturn or cutoff at the high end of the mass function \citep[e.g.,][]{larsen09,adamo15,johnson17,messa18,mok19,adamo20}. This upper mass cutoff is often modelled by a Schechter function, $dN/dM \propto M^{\beta} \mbox{exp}(-M/M_*)$, where $M_*$ is the cutoff mass. Here, we are interested in addressing two questions: What is the power-law index $\beta$ for young clusters in NGC~3351 and is there evidence for an upper mass cutoff?

We perform a least squares fit of the form: log~$dN/dM = \beta~\mbox{log}M + \mbox{const}$ to the distributions of the visually-classified class~1 and 2 clusters of NGC~3351 which is shown in Figure~\ref{fig:mass_function} as the red, dashed line and recorded in each panel. The results for $\beta$ range from about $-1.4$ to $-2.2$, consistent with the range found for cluster populations in other nearby galaxies \citep[e.g.,][]{fall12}, with the possible tendency to be somewhat shallower at the youngest ages. We also find a best fit value of $\beta=-1.75 \pm 0.23$ for clusters in the age interval of $1{-}200$~Myr, which was used by LEGUS \citep{adamo17}. 

We also perform maximum likelihood fits of the Schechter function to determine if the cluster mass functions are better fit by a Schechter function than a power-law, i.e., show statistically significant evidence for an upper mass cutoff. These fits follow the methodology described in \cite{mok19}, and have the advantage of not using binned or cumulative distributions. In Figure~\ref{fig:ML}, we plot the resulting $1\sigma$, $2\sigma$, and $3\sigma$ contours for the $M_*$ and $\beta$ from our maximum likelihood fits for the same three age intervals as before. For the two older age bins, the contours, including that at $1\sigma$, remain open all the way up to the maximum tested mass at the right edge of the plots. This indicates that no Schechter-like cutoff mass is detected, and the best-fitting value of $M_*$ is simply a lower limit. In the youngest age bin (ages between $1{-}10$~Myr), the 1$\sigma$ and 2$\sigma$ contours are closed but the $3\sigma$ contour remains open to the mass limit. In this case, there is weak evidence, at the $\sim$2$\sigma$ level, for a possible upper mass cutoff. However, since such a cutoff is not seen in the older cluster populations, we consider it unlikely that this is a physical feature present in the mass function of the NGC~3351 clusters.

\subsection{Stochasticity in Sampling the IMF}
\label{sec:stoch}

\begin{figure}
    \centering
    \includegraphics[width=\columnwidth]{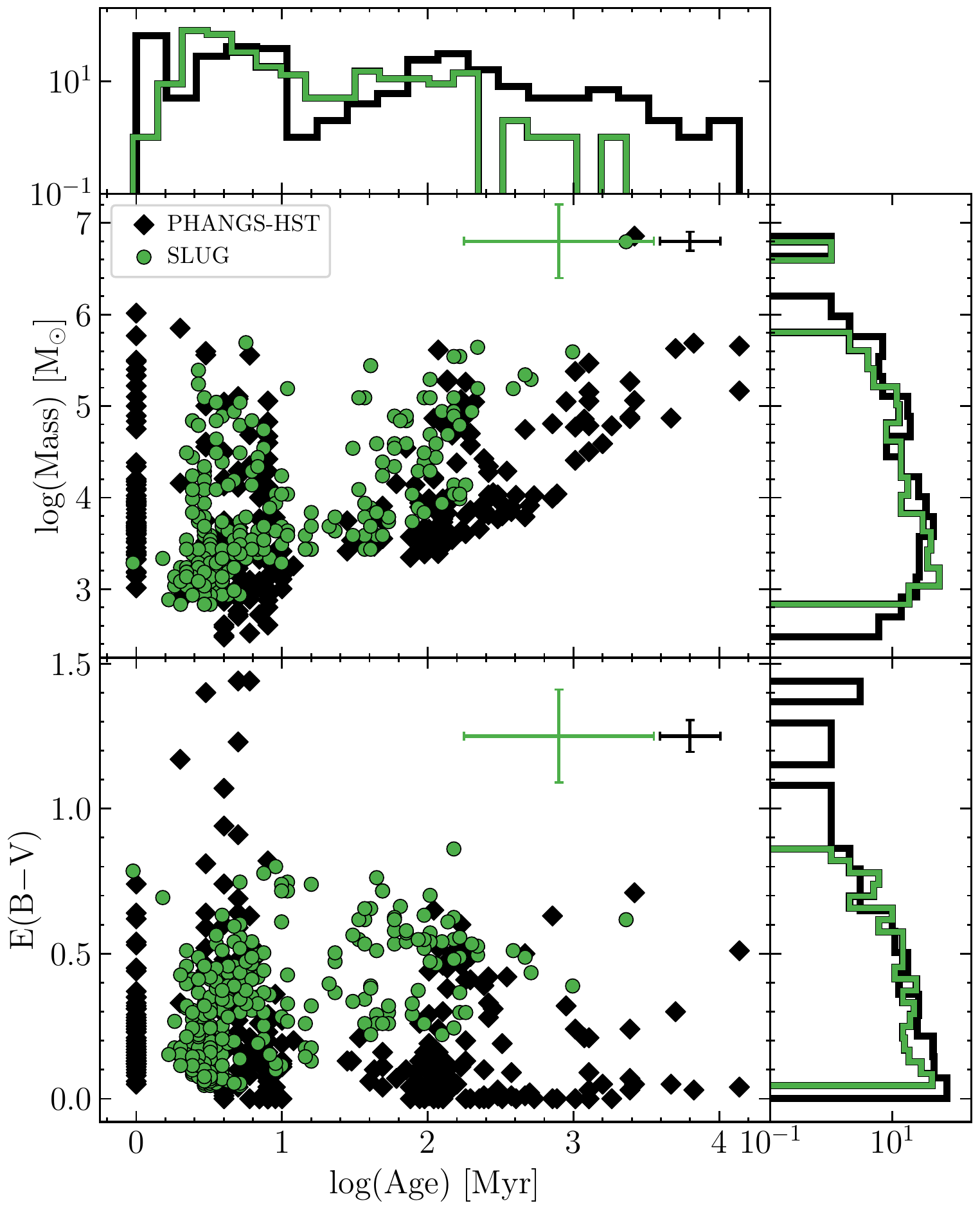}
    \caption{Comparison of the PHANGS--HST best-fitting results (black diamonds) to the \textsc{slug} results (green circles) for the class 1 and 2 clusters of NGC~3351. Logarithmic histograms on each axis show the distributions for both the PHANGS--HST and \textsc{slug} results. Typical error bars, computed from the median uncertainties on the properties, are shown in the corner of each panel. \textsc{slug} tends to find fewer old clusters and higher reddening values for the clusters older than 10~Myr. With \textsc{slug}, the majority of the clusters are found to be between 2 and 10~Myr and only a single cluster is found at 1~Myr. The mass distributions are similar, however, the PHANGS--HST results peak at slightly higher masses than \textsc{slug}.}
    \label{fig:slug}
\end{figure}

We use a deterministic approach to SED fitting which is justified for stellar populations that completely sample the stellar IMF. However, there is a low-mass cluster regime (<10$^{4}$~$M_{\odot}$) where the IMF may not be fully sampled. It becomes more likely that a low-mass stellar cluster will not contain high mass stars and the IMF will appear to be truncated. The stellar population will also be fainter than a higher-mass population which fully populates the IMF. This effect is called sampling stochasticity and it results in a broad range of possible luminosities and colors for these low mass clusters \citep{barbaro77,girardi93,lancon00,bruzual02,cervino06,deveikis08,fouesneau12,hannon19}. In color-color space, clusters, which may be stochastically sampling the IMF, can be found in the region bluer in $V{-}I$ than the SSP tracks at 1 to 5~Myr. These clusters have a deficiency of post-main sequence stars versus the fully populated IMF, i.e., the clusters appear bluer than expected from a lack of red post-main sequence stars \citep{fouesneau12}. On the other hand, when an excess of post-main sequence stars occurs, clusters are then found in the region redder than the SSP tracks. However, this is degenerate with reddening effects due to dust extinction, and can be lead to higher SED-fit reddening values \citep{hannon19}. 

Recently, work has been done to explore the modelling of stellar populations which do not fully sample the IMF. One such model is \textsc{slug}: Stochastically Lighting up Galaxies \citep{dasilva12,krumholz15}. As a sanity check, we run \textsc{cluster\_slug}\footnote{\url{http://www.slugsps.com/}} on our PHANGS--HST cluster catalog while including the fiducial prior function $p_{\rm prior} = M^{-2.0}T^{-0.5}$ used by \cite{krumholz15}. We use the \textsc{slug} star cluster model library \texttt{modp020\_chabrier\_MW} with a Milky Way extinction curve, Padova--AGB stellar evolution isochrones, solar metallicity, \textsc{starburst99 v7} stellar atmospheres, and the \cite{chabrier05} IMF. Figure~\ref{fig:slug} shows the comparison between the PHANGS--HST best-fitting results and the \textsc{slug} results. We find \textsc{slug} tends to characterize clusters as slightly less massive, younger, and with larger reddenings than our PHANGS--HST results. \textsc{slug} does not return any old clusters with little reddening which is in direct contrast with our results where we find a majority of the clusters beyond 100~Myr have low reddening values at or close to 0~mag. For our sample, the largest changes in cluster ages (from older to younger when using \textsc{slug}) occur in clusters with masses of less than about 10$^{4}$~$M_{\odot}$ which makes up 60 per cent of our cluster sample. The stochastic sampling of the IMF will have the largest impact when the youngest, most massive stars are missing. Since \textsc{slug} accounts for this, it naturally finds more lower-mass and younger star clusters.

\subsection{Application of Bayesian Priors}
\label{sec:prior}

In the previous sections, we have focused on the $\chi^{2}$ minimized SED fitting results for the clusters of NGC~3351. In the following analysis, we explore the application of astrophysically-motivated Bayesian priors to infer the star cluster properties. \textsc{cigale} generates a $\chi^{2}$ value (and subsequent likelihood value equal to $\exp(-\chi^{2}/2)$) for each model on the model grid. The priors scale the likelihoods of each model thus reshaping the PDF. \cite{krumholz15} tested priors of the form 
\begin{equation}
    \label{eq:prior}
    p_{\rm prior} \propto M^{\beta} T^{\gamma},
\end{equation}
where $M$ is the model mass and $T$ is the model age. The priors are a reflection of the expected observed clusters' mass and age distributions. Young star cluster observations have consistently found a mass distribution of $dN / dM \propto M^{-2}$ giving a $\beta$ value of $-2$ \citep[e.g.,][]{williams97,zhang99,bik03,degrijs03b,bastian11,fall12,fouesneau12,adamo17}. We find a consistent $\beta$ value for our sample of star clusters (Section~\ref{subsec:mf}). \cite{krumholz15} tested three values for $\gamma$: $\gamma = 0$ corresponding to a distribution that is flat in linear age; $\gamma = -1.0$ corresponding to a distribution flat in log(age) i.e., 90 per cent decline each decade of linear time \citep[e.g.,][]{whitmore07}; $\gamma = -0.5$ fiducial prior acting as a compromise and is most similar to observations for spiral galaxies. In this section, we test these values of $\beta$ and $\gamma$ on our sample. 

We first show how the priors are implemented in the analysis and how the cluster properties are inferred. The priors scale the likelihoods which gives the posterior probability distribution
\begin{equation}
    \label{eq:post}
    p_{\rm post}(M, T, A_{V}) \propto M^{\beta} T^{\gamma} \exp{\left(-\chi^{2}/2\right)}
\end{equation}
where $A_{V}=3.1/E(B{-}V)$. To derive a Bayesian estimate for a cluster's age, mass, and extinction, we calculate the expectation value of the posterior probability distribution. The expectation value of, for example, a cluster's age is defined as
\begin{equation}
    \label{eq:expect}
    \langle T \rangle = \frac{1}{C} \int dM \int dT \int dA_{V} \; T \; p_{\rm post}(M, T, A_{V})
\end{equation}
where
\begin{equation}
    \label{eq:c1}
    C = \int dM \int dT \int dA_{V} \; p_{\rm post}(M, T, A_{V}).
\end{equation}
However, given the discrete nature of the model grid in $M, T, A_{V}$ space, a quadrature approximation is calculated to give the expectation value
\begin{equation}
    \label{eq:approx}
    \begin{split}
        \langle T \rangle = &\frac{1}{C} \sum_{i} (M_{i+1} - M_{i}) \sum_{j} (T_{j+1} - T_{j}) \sum_{k} (A_{V k+1} - A_{V k})\\
        &T_{j} \; p_{\rm post}(M_{i}, T_{j}, A_{V k})
    \end{split}
\end{equation}
where
\begin{equation}
    \label{eq:c2}
    \begin{split}
        C = &\sum_{i} (M_{i+1} - M_{i}) \sum_{j} (T_{j+1} - T_{j}) \sum_{k} (A_{V k+1} - A_{V k})\\
        &p_{\rm post}(M_{i}, T_{j}, A_{V k}).
    \end{split}
\end{equation}
We then calculate the dispersion as the square root of the second moment of the posterior probability distribution:
\begin{equation}
    \label{eq:sigma}
    \sigma^{2}_{T} = \sum_{j}\left(T_{j} - \langle T \rangle\right)^{2} \; p_{\rm post}.
\end{equation}
However, this dispersion can be a poor estimation of the error on the Bayesian parameter estimate in cases with long-tailed or bimodal posterior PDFs. For such cases, the 16th to 84th percentile range is a more robust estimator of the error. For this paper, we will continue to use the dispersion as the uncertainty estimator but the 16th and 84th percentiles, along with the dispersion, will be included in the PHANGS--HST cluster catalogs. All calculations are performed in linear $M$ and $T$ space. The expectation value (Equation~\ref{eq:approx}) with corresponding uncertainty (Equation~\ref{eq:sigma}) acts as the Bayesian estimates for the cluster ages, masses, reddenings.

\begin{figure*}
    \centering
    \includegraphics[width=\textwidth]{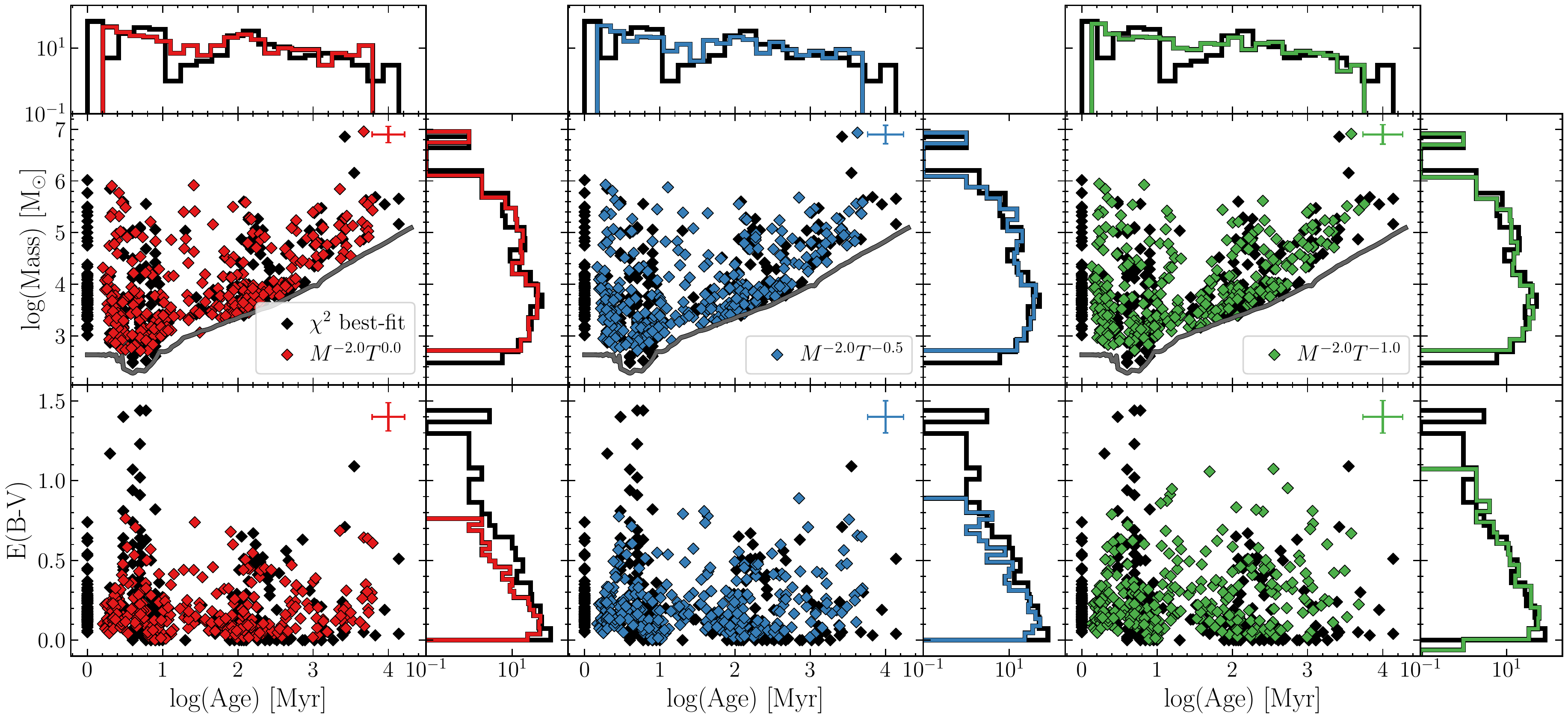}
    \caption{Bayesian estimated mass and reddening versus age for three different prior functions (blue, red, and green diamonds) compared to the $\chi^{2}$ best-fitting results (black diamonds) for PHANGS--HST photometry of star clusters in NGC~3351. The gray line in the upper panels indicates the magnitude limit $M_{V} = -6$ Vega mags. Logarithmic histograms beside each plot show the distribution of the ages, masses, and extinctions in each case. We adopt $p_{\rm prior} = M^{-2.0}T^{-0.5}$ (middle panels) as the fiducial prior. Typical error bars are computed as the median of the uncertainties and are given in corner of each panel. For the fiducial prior, the median uncertainties on the Bayesian age, mass, and reddening estimates are 0.24~dex, 0.18~dex, and 0.10~mag, respectively.}
    \label{fig:prior_compare} 
\end{figure*}

A comparison of the Bayesian parameter estimates is given in Figure~\ref{fig:prior_compare}. For all three priors tested, $\beta$ is set to $-2$. The steeper distribution with $\gamma = -1.0$ (green diamonds) shows fewer older clusters than the fiducial results (blue diamonds), which is to be expected. With $\gamma = 0.0$, clusters are found to extend to older ages which is, again, to be expected. Without sufficient physically-motivated criteria to evaluate which output age and reddening distribution depending on the chosen $\gamma$ value is more likely than another, it is difficult to definitively decide on the `best' prior. Instead, we quantify the differences in the estimated physical properties compared to the fiducial prior. With $\gamma = 0.0$, we find the median difference in cluster ages and masses to be $-0.048$~dex and $-0.005$~dex, respectively. The median difference in reddening is 0.014~mags. A negative difference in ages and a positive difference in reddening confirms that the $\gamma = 0.0$ prior returns older cluster ages with less reddening. With $\gamma = -1.0$, the median difference in cluster ages and masses is 0.057~dex and 0.003~dex, respectively. The median difference in reddening is $-0.017$ mags. A positive difference in ages and a negative difference in reddening confirms that the $\gamma = -1.0$ prior returns younger clusters with more reddening. In both cases, the cluster properties are fairly robust to the choice of prior. Given these results, we adopt the fiducial prior $p_{\rm prior} = M^{-2.0}T^{-0.5}$ using the method outlined above in order to compute the Bayesian inferred cluster property estimates. The quantified differences stated here can be treated as systematic uncertainties within the Bayesian estimates.

\begin{figure*}
    \centering
    \includegraphics[width=\columnwidth]{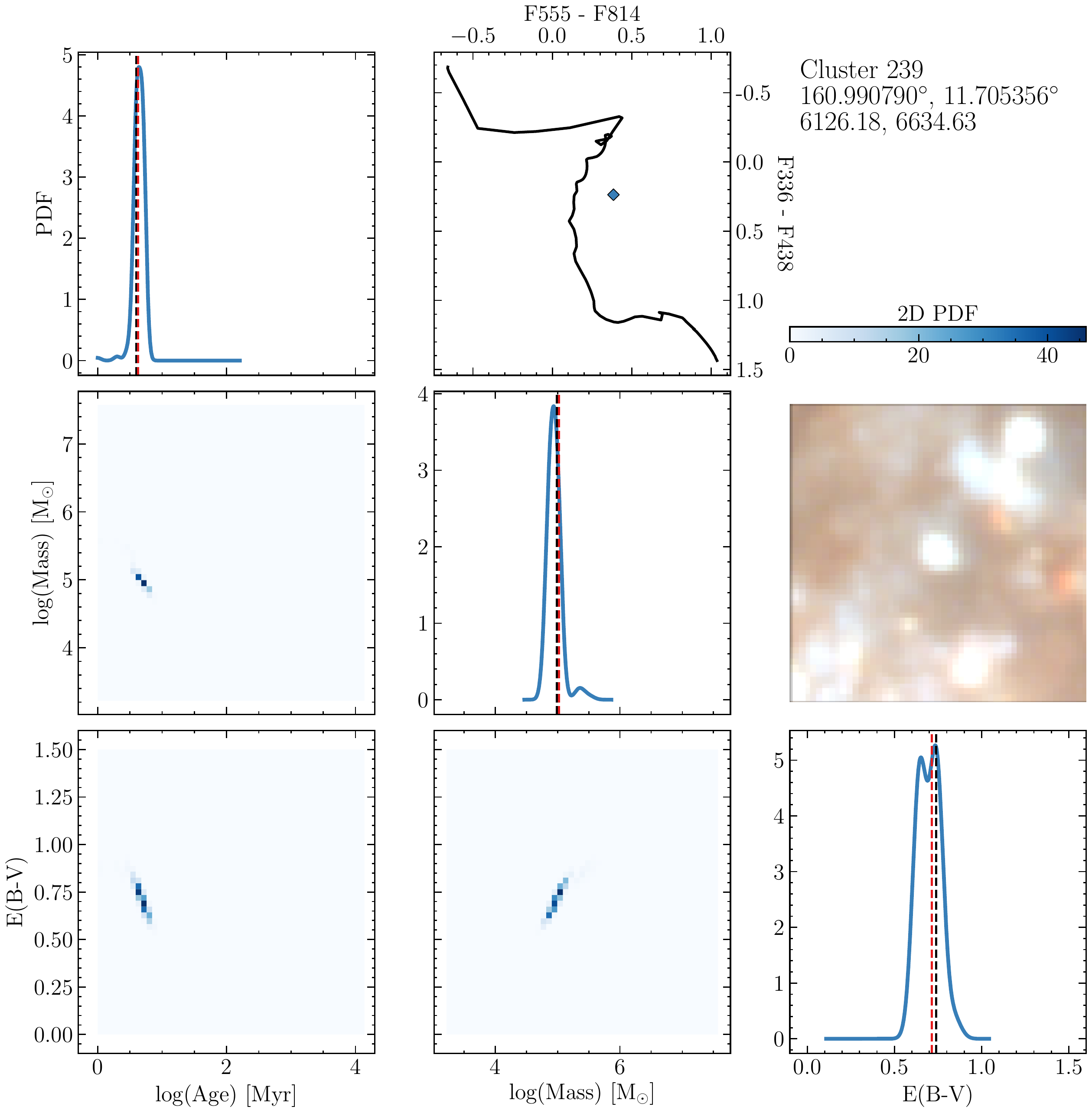} \hfill
    \includegraphics[width=\columnwidth]{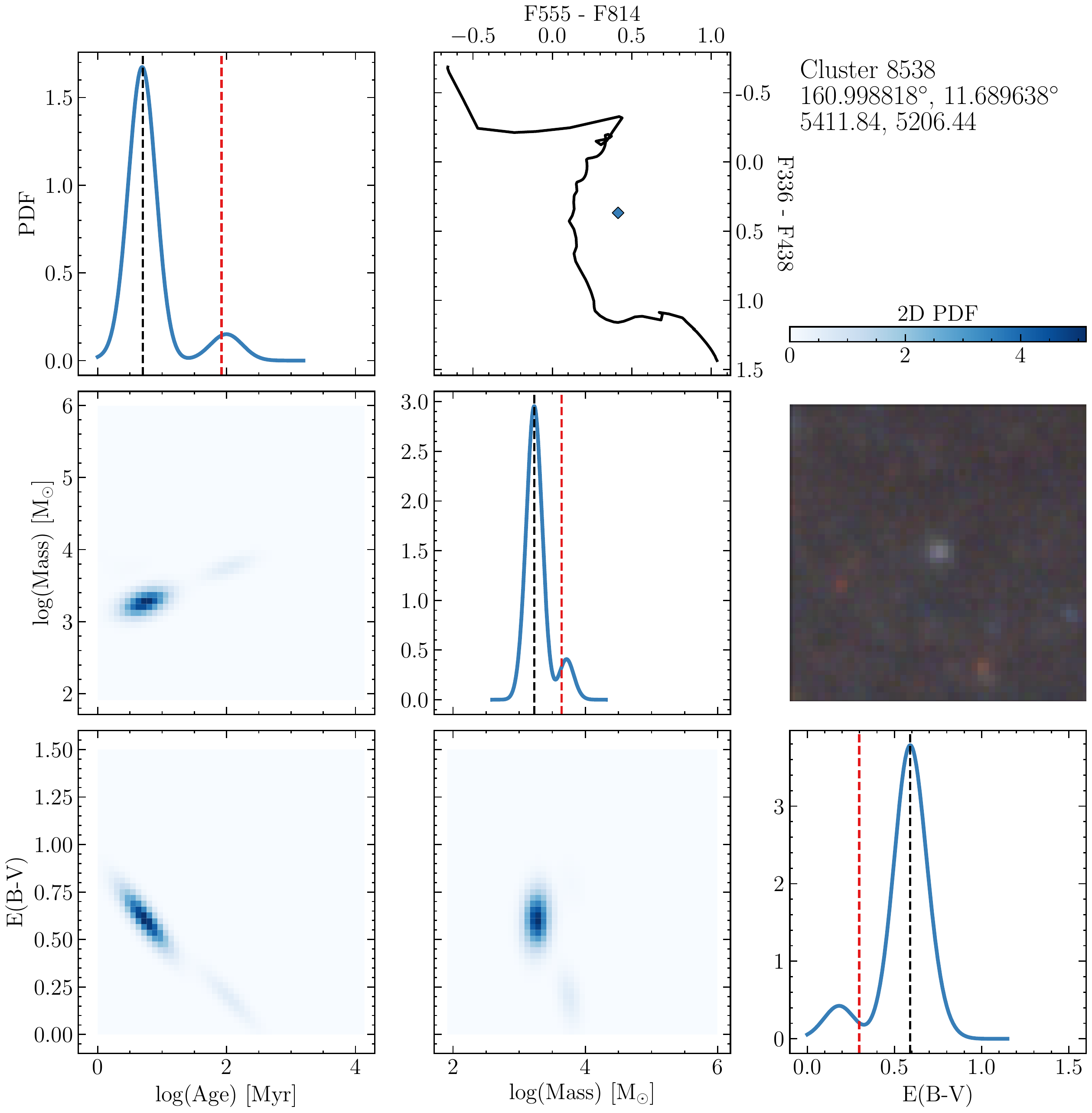}
    \caption{Corner plots of two example star clusters showing the age, mass, and reddening 1D and 2D kernel density estimated posterior PDFs after the application of the Bayesian fiducial prior. The black dashed lines are the minimized $\chi^{2}$ best-fitting result and the red dashed lines are the Bayesian estimates. The top-middle panel of each corner plot is a color-color diagram marking the location of the cluster and the \citetalias{bruzual03} model track. A postage stamp 3-color image of the cluster is given in the middle-right panel. Cluster~239 on the left is an example of a single-peaked age PDF with good agreement between the $\chi^{2}$ result and the Bayesian estimate. Cluster~8538 on the right shows a bimodal case where the Bayesian estimate and the $\chi^{2}$ result disagree.}
    \label{fig:corner}
\end{figure*}

Figure~\ref{fig:corner} shows example corner plots produced during the Bayesian analysis with the fiducial prior applied. For the corner plots, the posterior PDFs are computed by Gaussian kernel density estimation (which is only used for a visual representation of the PDFs and has no bearing on the computation of the expectation values). In the first case, cluster~239, the PDFs show singly-peaked distributions indicating a well-constrained Bayesian estimate. This case also demonstrates good agreement between the minimized $\chi^{2}$ results (black dashed lines) and the Bayesian estimates (red dashed lines). The second case, cluster~8538, demonstrates the age-reddening degeneracy where two cases are likely: a young, reddened cluster or an old cluster with little reddening. This is a case where Bayesian estimate disagrees with the $\chi^{2}$ result. We inspect each cluster's age and reddening PDFs and find $\sim$70 clusters with bimodality before the application of the Bayesian prior. After the application of the Bayesian prior, we find $\sim$30 bimodal cases ($\sim$10 per cent of the entire cluster sample) when inspecting the newly modified PDFs. The Bayesian analysis proves to be advantageous to help break degeneracies but not for all of the cases (as with cluster~8538). More information is needed to further reduce the number of bimodal cases. One option is to visually inspect each cluster in a 3-color image to look for obvious signs (to humans) of dust extinction. However, this is not feasible for the automated PHANGS--HST pipeline of thousands of star clusters without a well-taught and tested machine learning algorithm. In Section~\ref{sec:future}, we discuss other possible solutions for breaking the age-reddening degeneracy.

In a similar vein, in this paper, we are able to identify bimodal cases simply by inspecting each cluster's corner plot. However, for the full, automated PHANGS--HST pipeline, this will not be possible. The bimodal cases generally find disagreement between the $\chi^{2}$ minimized result and the Bayesian estimate so clusters with large disagreements could be flagged as tentatively bimodal within the pipeline. For these cases, instead of calculating the expectation value, the PDFs could be split between the two modes and new Bayesian estimates could be computed for each of the modes. Before this can be implemented, we will need to decide how to consistently split apart the two peaks in the PDFs and how the Bayesian estimates from the two modes are handled with regards to the final cluster properties used in the cluster catalogs. Other methods could be used to extract a Bayesian estimate from the PDF such as taking the median with 16th and 84th percentiles or finding the model with the maximum likelihood (i.e., the peak of the posterior PDF). 

\begin{figure}
    \centering
    \includegraphics[width=\columnwidth]{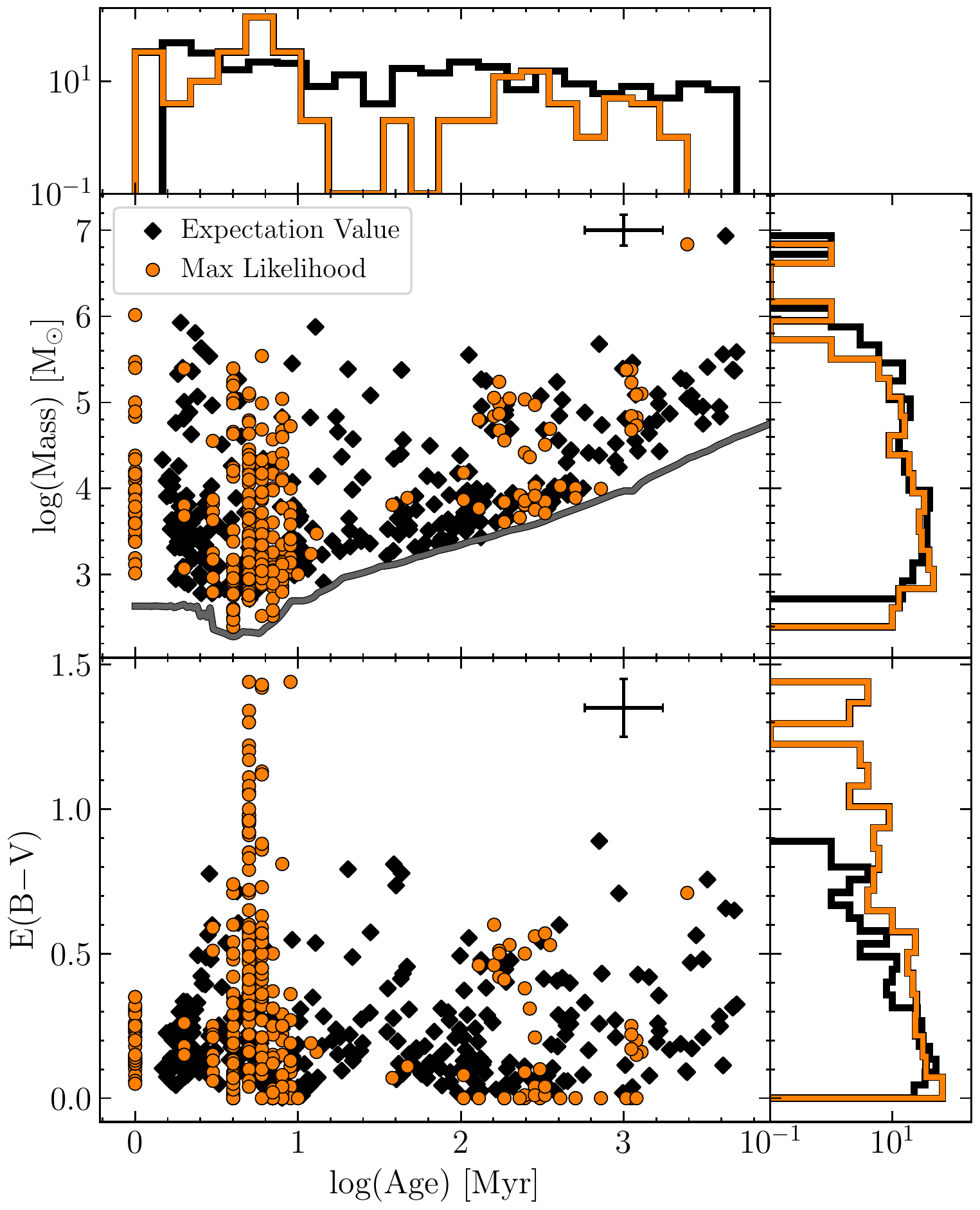}
    \caption{Bayesian estimates from the expectation values (black diamonds) compared to maximum likelihood Bayesian estimates (orange circles) assuming the fiducial prior. Logarithmic histograms show the distributions of the cluster properties for the two cases along each axis. The expectation value method finds a large dispersion in reddening for the clusters with ages from 10~Myr to 100~Myr while the maximum likelihood method only finds two clusters in this age range; the clusters are shifted to young ages which leads to the large spike in reddening at the young ages. The maximum likelihood method is able to find clusters at 1~Myr which is not found with the expectation value method. The mass distribution is robust between the two Bayesian estimate methods.}
    \label{fig:maxlike}
\end{figure}

Figure~\ref{fig:maxlike} compares the cluster properties as derived from the expectation value and maximum likelihood methods. The expectation value results reveal a few interesting, and possibly worrisome, trends. The method fails to find very young clusters (youngest cluster is 1.5~Myr) due to 1~Myr being the boundary. All the available models are older than 1~Myr which skews the expectation value (a weighted average) to be older than 1~Myr. We find a slight trend for older clusters to have larger reddenings which is counter-intuitive. This is in contrast to the negative correlation between age and extinction found by \cite{bastian05} for the star clusters of M51. However, \cite{grasha18} find a poor correlation between star cluster age and extinction in NGC~7793 based on SED-fitting from the LEGUS program. Further, \cite{grasha18} find that star clusters not associated with any GMCs have a slightly higher extinction than those still semi-embedded in their natal gas clouds, which may be unexpected. The same result is found for the star clusters of M51 \citep{grasha19}. This serves as a testament to the complexities of star cluster SED modelling and the degeneracy of age and reddening. We see a maximum reddening of around 0.9~mag with a low dispersion in reddening for the youngest clusters and a larger dispersion in reddening for clusters from 10~Myr to 100~Myr. The majority of the clusters in this age range appear to be the ones identified as bimodal which makes sense given the high probability of age-reddening degeneracy at this location of the SSP model track (Figure~\ref{fig:color_color_tracks}).

The maximum likelihood results show a few contrasting trends. We see a quantization of cluster ages, especially at 10~Myr and younger. This is to be expected as the models available for the method to choose from are those following the model grid. We do find clusters at 1~Myr using this method. We find a significant portion of clusters at ages from 5 to 9~Myr with a virtual lack of clusters between 10~Myr to 100~Myr. The clusters found to be around 20~Myr with the expectation value method appear to be shifting all to ages younger than 10~Myr with the maximum likelihood method. We also find roughly half of the clusters older than 100~Myr to have very low and even zero reddenings. We check if these are clusters identified as globular cluster candidates and find only one of them to be a globular cluster candidate. The remaining globular cluster candidates are found in the grouping at around 1~Gyr with 0.2~mag reddening, a few at younger ages around 200~Myr with 0.5~mag reddening, and, most troubling, 30 per cent of the globular clusters are found at 5~Myr old. In contrast, the expectation value method finds the youngest globular cluster candidate at 130~Myr and the next youngest at 180~Myr. The remaining ones are at 200~Myr up to 4~Gyr with an average reddening of 0.33~mag.

While the expectation value method returns possibly unreliable results for the bimodal cases (about 10 per cent of our NGC~3351 sample), the maximum likelihood method is not a perfect solution. By choosing the single model with the highest likelihood, information within the PDF is ignored. The maximum likelihood model may be surrounded by unlikely models while the second mode of the PDF, at a lower relative likelihood, could be encompassed by equally likely models which gives better confidence that the second mode may actually provide a better estimate of the cluster's physical properties. This also makes it difficult to characterize the uncertainties on the maximum likelihood result. With the expectation value, the uncertainties on the estimate can be calculated as the square root of the second moment of the posterior PDF which describes the width of the peak in the PDF and therefore the uncertainty of the estimate. By ignoring the information available in the PDF, there is no way to measure the uncertainty on the maximum likelihood result. 

More work is needed to adequately explore these trends and effects before the Bayesian analysis can be fully implemented into the PHANGS--HST SED fitting pipeline. We have examined the expectation value as the Bayesian estimate but have highlighted potential issues with these estimates, particular for those objects with multimodal PDFs, in this section. In the following section, we discuss future efforts with the goal to resolve the outstanding issues with the Bayesian analysis, the age-reddening degeneracy, and other areas of the SED fitting which need further exploration.

\section{Future Work}
\label{sec:future}

There are still a number of tweaks that can be made to the SED modelling procedures discussed here with the goal of providing the most accurate and robust star cluster properties across the PHANGS--HST sample as possible. Here we outline future work that can be done to help reach this goal.

One of the biggest obstacles for all star cluster SED fitting endeavours is the age-reddening degeneracy. The answer usually involves the incorporation of more data and how to best apply those data. For the PHANGS--HST sample, one option is to supplement the cluster photometry with high-resolution H$\alpha$ observations. \cite{whitmore20} detail methodology for using \textit{HST} H$\alpha$ observations to improve upon the ages provided by star cluster SED fitting. This could be applied to our PHANGS--HST pipeline if such observations are available for our sample. Currently, there are less than a dozen galaxies within the PHANGS--HST sample that have \textit{HST} H$\alpha$ imaging, and archival observations do not necessarily cover the same footprint as PHANGS--HST as is the case for NGC~3351. Ground-based H$\alpha$ could in principle be used, but the impacts of their significantly coarser resolution ($\sim$1\arcsec) would first need to be understood. The PHANGS collaboration has PHANGS--MUSE $\sim$1$\arcsec$ H$\alpha$ maps available for 19 of the PHANGS--HST galaxy sample. Ground-based H$\alpha$ maps from narrowband photometry have been compiled by the PHANGS collaboration (A.\ Razza et al. in prep.) which cover all PHANGS galaxies. 

One additional option to potentially break the degeneracy is to apply a different prior for the reddening $E(B{-}V)$. Currently, we assume a flat prior where all reddening models are equally likely. However, a non-flat prior is an option, particularly a log-normal distribution on the basis that the gas column density PDF, which is completely decoupled from star cluster work, follows a log-normal distribution. With LEGUS star clusters, \cite{ashworth17} produce an observed extinction distribution that strongly disfavors high $A_V$. This result indicates that a non-flat prior for the reddening could be necessary.

Another possibility for breaking the age-reddening degeneracy is to take advantage of the PHANGS high-resolution ALMA CO maps. Using a gas-to-dust ratio to convert the CO maps into dust maps, we could match the star clusters with the dust map to identify how much dust is present at the cluster location. This information would then inform which mode of the bimodal age-reddening distribution to choose as the best estimate for the age and reddening. However, there are line-of-sight uncertainties with this simple matching. It is unknown whether the cluster resides in front of, within, or behind the dust. Regardless, a lack of CO emission would provide a strong prior that there is little extinction. Additionally, the gas-to-dust conversion introduces uncertainties and similar to the ground-based H$\alpha$ data, there is also a mismatch in resolution, and the impacts of which would need to be examined since the CO maps have resolutions of $\sim$1\arcsec. Despite these drawbacks, the ALMA CO maps are still a promising direction in which to pursue, as they are, by design, available for the full PHANGS--HST sample. In conjunction with this method, we can identify dust lanes in the PHANGS--HST optical imaging observations and apply this dust information to the SED fitting as well. 

As discussed in Sections~\ref{sec:dust}, PHANGS--MUSE observations produce Balmer decrement maps as well as stellar $E(B{-}V)$ maps from stellar continuum fitting at resolutions similar to ALMA (1$\arcsec$). \cite{pellegrini20b} find Balmer decrement measurements work well for de-reddening H$\alpha$ fluxes in model galaxies using the population synthesis model \textsc{warpfield}. This is a promising result which supports the use of PHANGS--MUSE Balmer decrement measurements to inform our SED modelling. 

\begin{figure*}
    \centering
    \includegraphics[width=\textwidth]{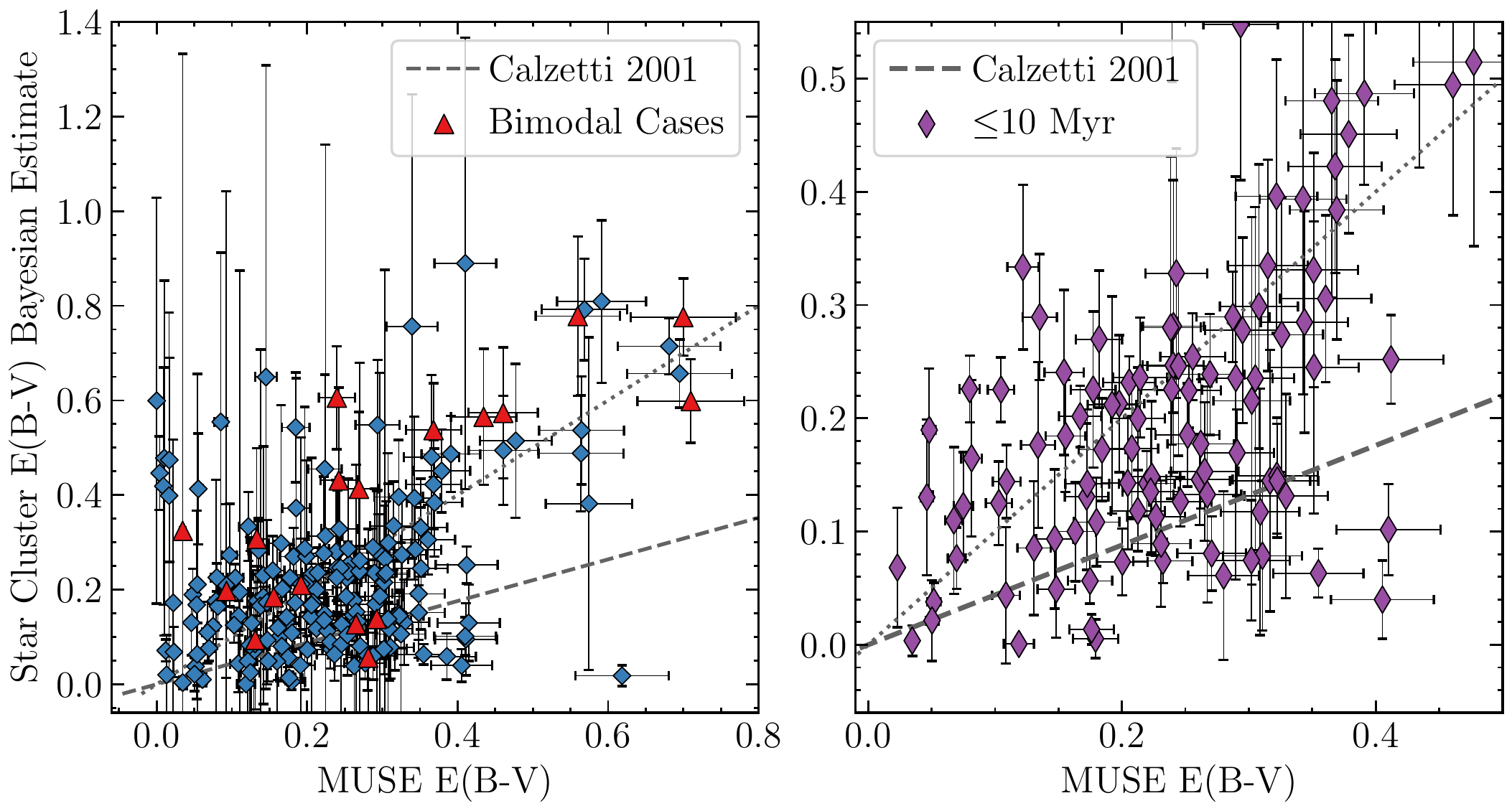}
    \caption{Star cluster Bayesian estimates for $E(B-V)$ versus the MUSE $E(B-V)$ measurements at the location the star clusters. MUSE $E(B-V)$ measurements are derived from Balmer decrements to calculate $A_{V}$ assuming $R_{V} = 4.05$. 266 clusters lie within the PHANGS--MUSE footprint but 75 clusters have no Balmer decrement measurement at their location. Clusters with bimodal PDFs are highlighted as red triangles. The dotted line shows unity and the dashed line marks the factor of $1/0.44$ times greater reddening of gas compared to stars \citep[][equation 9]{calzetti01}. $1\sigma$ error bars are given for the Bayesian estimates and 10 per cent error bars are given for the MUSE reddening values. The right panel shows a subset of the left panel highlighting only the clusters which are found to be 10~Myr and younger.}
    \label{fig:muse_bimodal}
\end{figure*}

Here, as a first step, we can check if the Bayesian analysis from Section~\ref{sec:prior} is inferring reddening values that are in agreement with $E(B{-}V)$ values from the PHANGS--MUSE data. The PHANGS--MUSE $E(B{-}V)$ measurements are derived from $A_{V}$ maps from Balmer decrement measurements. We match the location of each star cluster to the closest pixel in the PHANGS--MUSE map and measure the mean $E(B{-}V)$ value of a three by three pixel grid (0.36~arcsec$^{2}$) centered on that pixel. We compare our star cluster Bayesian estimates to the MUSE measurements in Figure~\ref{fig:muse_bimodal} and find a weak correlation between the two $E(B{-}V)$ measurements. In the left panel of Figure~\ref{fig:muse_bimodal}, 18 of the bimodal cases which have MUSE Balmer decrements at their locations are highlighted. The majority of bimodal cases agree with the MUSE measurements within uncertainties. A few of the bimodal cases could possibly be resolved by including the information provided by the MUSE data. More of the star clusters not identified as bimodal could possibly be improved by using the MUSE measurements. However, more attention beyond the scope of this paper will need to be devoted to such analysis before it can be successfully integrated into the PHANGS--HST star cluster catalog pipeline.

The right panel of Figure~\ref{fig:muse_bimodal} focuses on a potentially encouraging result. All the clusters with Bayesian estimates of 10~Myr and younger are plotted along with the relation $E(B{-}V)_{\rm star} = 0.44 E(B{-}V)_{\rm gas}$ \citep[][equation 9]{calzetti01} which gives roughly a factor of 2 greater reddening in gas compared to stars (or star clusters in our case). Clusters at these young ages should still have H$\alpha$ associated with them and the MUSE reddening measurements (derived from Balmer decrements) depend on H$\alpha$ detection. We see a potential correlation along the \citet{calzetti01} line for some of the young star clusters which may explain the deviation from unity for the clusters. Thus, it is encouraging that our star cluster reddening estimates are consistent (within the large scatter) with the values from MUSE for these young clusters along the expected relation.

In Section~\ref{sec:ssp}, we discuss the choice to not include a nebular emission component in our SED fitting. However, there is a case to be made to include nebular continuum and line emission for clusters that lie bluer than the SSP tracks at 1~Myr to 5~Myr in color-color space (see Figure~\ref{fig:color_color_tracks}). In our NGC~3351 star cluster sample, we find three clusters in this region. For such clusters, including a nebular emission component in the fitting could be justified. Additionally, other clusters could intrinsically have colors in this same region but have been reddened by dust and therefore lie to the right of the SSP track in color-color space. This approach will need to be tested to see if it makes a significant difference in the SED fitting results, since the solar metallicity tracks at those ages are very similar, to warrant implementing it into our pipeline in the future.

In Section~\ref{sec:prior}, we discuss the application of Bayesian priors to modify the likelihood values of each model on the grid. A final Bayesian parameter estimate is calculated as the expectation value or finding the model with the maximum likelihood. We show these methods only work well for certain populations and fail for others (e.g., bimodal cases). Going beyond these simple methods, we could incorporate each cluster's entire PDF into computing the mass and age functions of the cluster population. This will circumvent the need to fix every single bimodal case while utilizing the information-rich PDFs. 

\section{Conclusions}
\label{sec:conclusions}
The PHANGS--HST project aims to study individual star clusters in 38 nearby galaxies making use of the publicly available SED fitting code \textsc{cigale} to characterize their physical properties. 

\begin{enumerate}
    \item We test \textsc{cigale}'s ability to recover known cluster properties from a mock catalog within the photometric uncertainties and find good recovery of ages (standard deviation of the difference between the `true' and recovered ages of 0.31~dex), masses (standard deviation of the difference of 0.18~dex), and reddenings (standard deviation of the difference of 0.09~mag). The largest age residuals are found at 1~Myr, at around 10~Myr, and at the very oldest ages. Degeneracies in the SEDs of the SSP models at 5~Myr to 50~Myr are the major cause of the large age residuals. We find no change in this result when using finer or coarser age grid sampling. We benchmark \textsc{cigale}'s ability to recover the same SED fitting results as the LEGUS stellar cluster catalogs and can successfully recover the LEGUS results. The median value of the ratios between the \textsc{cigale} and LEGUS ages is $0.001 \pm 0.017$~dex and the median of the mass ratios is $0.003 \pm 0.011$~dex.
    \item Using the visually-classified class~1 and 2 subset of the PHANGS--HST star cluster catalog for NGC~3351, we test the SED modelling dependencies. We consider the differences between two single stellar population models (\cite{bruzual03} and \textsc{yggdrasil}) and the inclusion of nebular emission within those models. We explore how to treat extinction and reddening from dust in the SED modelling and evaluate our chosen modelling parameters against IFU data from PHANGS--MUSE. 
    \item We test how fitting in linear fluxes or logarithmic magnitudes affects the SED results and find no significant changes in the resulting distributions of the ages, masses, and reddenings. 
    \item Based on the results of our tests in this paper, we choose to adopt for our SED fitting: linear fluxes, the \citetalias{bruzual03} SSP model, solar metallicity, an instantaneous burst star formation history, a fully sampled \citet{chabrier03} IMF, intrinsic reddening varying from 0 to 1.5~mag in 0.01~mag steps, and no nebular emission component. We sample age models linearly in 1~Myr intervals for 1~Myr to 10~Myr and logarithmically for 11~Myr to 13.75~Gyr ($\Delta\log({\rm Age/Myr}) \approx 0.3$). This gives an age-reddening model grid with 16,610 models. 
    \item We apply this SED modelling approach to derive the best-fitting star cluster ages, masses, and reddenings for NGC~3351. We find that the star clusters within the inner ring at the center of NGC~3351 to be young and more massive than the rest of the clusters, which may be due to the funneling of gas into the center by the galactic bar. This funnelling triggers and maintains ongoing star formation.
    \item The clusters present in the stellar bulge region are found to be much older on average, consistent with expectations from previous work.
    \item The SED fitting results are checked against visually-identified globular cluster candidates and, although they are found to be older relative than the rest of the cluster sample, the ages fall short of the expected age of 10~Gyr. This is most likely due to the solar-metallicity SSP model being used for an old, low-metallicity cluster population as well as the preference for the SED fitting to choose younger, higher reddened models \citep[see][]{whitmore20}. Thus, caution should be exercised when studying globular cluster ages derived from broad-band SED fitting. Clusters with visually-identified HII regions are also checked, and the SED ages are below 10~Myr as is expected. 
    \item We study the cluster mass functions and find power-law slopes of $\beta \sim -2$, consistent with the literature. Maximum likelihood fits find no evidence of an upper mass cutoff for the clusters of NGC~3351 down to our completeness limit at $10^{3.5}$~$M_{\odot}$ for the $10{-}100$~Myr age bin and $10^{4.8}$~$M_{\odot}$ for the $100{-}400$~Myr age bin.
    \item We explore the application a Bayesian prior of the form $p_{\rm prior} \propto M^{\beta} T^{\gamma}$ to modify the marginalized PDFs and derive Bayesian estimates for the cluster properties. Without sufficient physically-motivated criteria to evaluate which values of $\beta$ and $\gamma$ provide the most-likely output distributions, we choose to adopt $\beta = -2$ and $\gamma = -0.5$ as our fiducial prior, and quantify the differences from this prior. We find that our Bayesian estimates perform poorly for clusters at 1~Myr and clusters with bimodal PDFs. Further testing of the Bayesian analysis is needed to diagnose and mitigate these issues.
    \item We find agreement with large scatter between our star cluster reddening Bayesian estimates and the MUSE reddening measurements (based on Balmer decrements). For clusters 10~Myr and younger, the disagreement between our measurement and MUSE may be explained by the \citet{calzetti01} relation \mbox{$E(B{-}V)_{\rm star} = 0.44 E(B{-}V)_{\rm gas}$}.
\end{enumerate}

The SED fitting methodology detailed in this paper will be the basis for estimating star cluster physical properties in the forthcoming publicly available PHANGS--HST star cluster catalogs. The methodology will also be applied to the stellar associations identified in the PHANGS--HST sample which will be detailed in K.\ L.\ Larson et al. (in prep.). In short, a watershed algorithm is used to select the stellar associations of star clusters which provides a better way of identifying the youngest and least-massive clusters, and yields a more complete picture than detection of the difficult-to-model asymmetric class~3 clusters. Applying the same SED fitting methodology with possible modifications for more complex star formation histories will provide self-consistency across both the PHANGS--HST star cluster catalog and the PHANGS--HST stellar association catalog. 

The physical properties derived by SED fitting for the PHANGS--HST cluster and stellar association catalogs will be crucial for studying the connections between molecular clouds and young star clusters across diverse galactic environments. The catalogs will provide both the $\chi^{2}$ minimized best-fitting results, the Bayesian parameter estimates for star cluster ages, masses, and reddenings, and the associated uncertainties. These catalogs, including the catalog for NGC~3351 used in this paper, will be made publicly available through the Mikulski Archive for Space Telescopes (MAST).

\section*{Acknowledgements}

This work is based on observations made with the NASA/ESA Hubble Space Telescope, obtained at the Space Telescope Science Institute, which is operated by the Association of Universities for Research in Astronomy, Inc., under NASA contract NAS 5-26555. These observations are associated with program \#13364. This research has made use of the NASA/IPAC Extragalactic Database (NED) which is operated by the Jet Propulsion Laboratory, California Institute of Technology, under contract with NASA. Based on observations and archival data obtained with the \textit{Spitzer} Space Telescope, which is operated by the Jet Propulsion Laboratory, California Institute of Technology under a contract with NASA. Based on observations collected at the European Southern Observatory under ESO programs 1100.B-0651, 095.C-0473, and 094.C-0623. This work was carried out as part of the PHANGS collaboration. JCL acknowledges the W.M. Keck Institute for Space Studies (KISS) for its support of PHANGS-HST collaboration meetings where work for this paper was completed, and benefited from initial discussions at the 2014 KISS workshop, ``Bridging the Gap: Observations and Theory of Star Formation Meet on Large and Small Scales."  MB acknowledges partial support from FONDECYT regular 1170618. PSB acknowledges support through the RAVET project PID2019-107427GB-C31 from the Spanish Ministry of Science, Innovation and Universities. JMDK gratefully acknowledges funding from the German Research Foundation (DFG) in the form of an Emmy Noether Research Group (grant number KR4801/1-1) and from the European Research Council (ERC) under the European Union's Horizon 2020 research and innovation program via the ERC Starting Grant MUSTANG (grant agreement number 714907). KK gratefully acknowledges funding from the Deutsche Forschungsgemeinschaft (DFG, German Research Foundation) in the form of an Emmy Noether Research Group (grant number KR4598/2-1, PI Kreckel). R.S.K.\ acknowledges financial support from the DFG via the collaborative research center (SFB 881, Project-ID 138713538) ``The Milky Way System" (subprojects A1, B1, B2, and B8). He also thanks for subsidies from the Heidelberg Cluster of Excellence {\em STRUCTURES} in the framework of Germany's Excellence Strategy (grant EXC-2181/1 - 390900948) and for funding from the European Research Council (ERC) via the ERC Synergy Grant {\em ECOGAL} (grant 855130). FB acknowledges funding from the European Union’s Horizon 2020 research and innovation program (grant agreement No 726384/EMPIRE). ER acknowledges the support of the Natural Sciences and Engineering Research Council of Canada (NSERC), funding reference number RGPIN-2017-03987. ES and TGW acknowledge funding from the European Research Council (ERC) under the European Union’s Horizon 2020 research and innovation program (grant agreement No. 694343).

\section*{Data Availability}
The PHANGS--HST star cluster catalogs for NGC~3351 and the remaining PHANGS--HST galaxies will be made publicly available through the Mikulski Archive for Space Telescopes (MAST) in the near future. In the meantime, the NGC~3351 cluster catalog used in this article will be shared on reasonable request to the corresponding author.


\bibliographystyle{mnras}   
\bibliography{all}  



\bsp	
\label{lastpage}
\end{document}